\documentstyle[sprocl,epsf]{article}

\def\Journal#1#2#3#4{{#1} {\bf #2}, #3 (#4)}

\def\NPB{{\em Nucl. Phys.} B}
\def\PLB{{\em Phys. Lett.}  B}
\def\PRL{{\em Phys. Rev. Lett.}}
\def\PRD{{\em Phys. Rev.} D}

\def\JHEP{{\em J. High Energy Phys.}}

\def\mco{\multicolumn}

\def\ra{\rightarrow}

\def\ko{K^0}

\def\al{\alpha}

\def\be{\begin{equation}}
\def\ee{\end{equation}}
\def\bea{\begin{eqnarray}}
\def\eea{\end{eqnarray}}

\def\Lgr{{\cal L}}
\newcommand{\vev}[1]{\langle#1\rangle}
\def\be{\begin{equation}}
\newcommand{\bel}[1]{\begin{equation}\label{#1}}
\def\ee{\end{equation}}
\newcommand{\eref}[1]{(\ref{#1})}
\newcommand{\Eref}[1]{Eq.~(\ref{#1})}
\newcommand{\rem}[1]{}

\def\half{{1\over 2}}
\def\MM{{\cal M}}
\def\NN{{\cal N}}

\def\none{$\NN=1$}
\def\ntwo{$\NN=2$}

\def\nfour{$\NN=4$}
\def\susy{supersymmetry}
\def\susic{supersymmetric}

\def\rarr{\rightarrow}

\def\ZZ{{\bf Z}}
\def\ZN{\ZZ_N}

\def\NO{Nielsen-Olesen}
\def\ures{{\underline {\it Conclusion}:} }

\newcommand{\AmS}{{\protect\the\textfont2
  A\kern-.1667em\lower.5ex\hbox{M}\kern-.125RMS}}

\def\slash{\!\!\!\!/ \ }
\def\rarr{\rightarrow}

\def\NN{{\cal N}}
\def\order{{\cal O}}

\def\none{$\NN=1$}
\def\ntwo{$\NN=2$}

\def\nfour{$\NN=4$}
\def\susy{supersymmetry}
\def\susic{supersymmetric}
\def\Lgr{{\cal L}}
\def\gt{gauge theory}
\def\nc{N_c}
\def\nf{N_f}
\def\lam{\lambda}
\def\Lam{\Lambda}
\def\al{\alpha}
\def\mn{{\mu\nu}}

\def\ZZ{{\bf Z}}
\def\ZN{\ZZ_N}

\newcommand{\FFig}[1]{Fig.~{\ref{fig:#1}}}

\renewcommand{\thanks}{\footnote}
\def\tocite#1{$^{\hbox{\,-}}$\kern-.04em\cite{#1}}

\def\JLone<#1,#2>{#1}
\def\JLtwo<#1,#2,#3>{#2}
\def\JLyear<#1,#2,#3,#4>{#3}
\def\JLpage<#1,#2,#3,#4>{#4}

\def\Jpage<#1,#2,#3>{#3}

\begin{document}

\title{MILLENNIAL MESSAGES FOR QCD FROM THE SUPERWORLD AND FROM THE STRING}

\author{ M.J. STRASSLER }

\address{{Department of Physics, 
University of Washington, P.O.~Box 351560, 
Seattle WA 98195-1560 USA}}


\maketitle\abstracts{ Supersymmetric gauge theories have had a
significant impact on our understanding of QCD and of field theory in
general.  The phases of \none\ supersymmetric QCD (SQCD) are
discussed, and the possibility of similar phases in non-supersymmetric
QCD is emphasized.  It is described how duality in SQCD links many
previously known duality transformations that were thought to be
distinct, including Olive-Montonen duality of \nfour\ supersymmetric
gauge theory and quark-hadron duality in (S)QCD.  A connection between
Olive-Montonen duality and the confining strings of Yang-Mills theory
is explained, in which a picture of confinement via non-abelian
monopole condensation --- a generalized dual Meissner effect ---
emerges explicitly.  Similarities between supersymmetric and ordinary
QCD are discussed, as is a non-supersymmetric QCD-like ``orbifold'' of
\none\ Yang-Mills theory.  I briefly discuss the recent discovery that
gauge theories and string theories are more deeply connected than ever
previously realized.  Specific questions for lattice gauge theorists
to consider are raised in the context of the first two topics. 
(Published in ``At the Frontier of Particle Physics: Handbook of QCD'',
M. Shifman, editor)}

\section{Introduction: Beyond QCD}

The field of supersymmetric gauge theory is fascinating, technical,
and vast, and obviously cannot be reviewed in a single article.  My
limited intention is to give a qualitative and conceptual overview,
with an eye toward conveying what I see as some of the most important
messages of recent advances for our understanding of
QCD. Consequently, this article will consist largely of results that I
see as relevant to QCD; only rarely will it present the evidence for
these results or many technical details, and referencing will be
limited. The interested reader is directed to more focussed reviews on
the superworld.\cite{phasesreview,Bilal,CCM,AdSreview}

The purpose of my writing such a review?  This is perhaps conveyed best
by the words of Tom DeGrand, when he once asked me to speak to an
audience of lattice gauge theorists: he requested that I encourage
them ``to change the line in their code that sets $N_c,N_f=3$.'' I
hope to convince you that a systematic exploration of theories {\it
other} than real-world QCD is an important and exciting direction for
research, and that placing QCD in the context of a wider variety of
theories may become a powerful tool for understanding it.

There are several reasons why physicists should seek a detailed
understanding of gauge theories beyond real-world QCD.  First, we can
use them to gain insight into the properties of the real world.  It
would be helpful to know which aspects of the strong interactions are
special to $\nc=\nf=3$ and which ones are generic, or at least common
to many models. Second, a theory with behavior only vaguely similar to
QCD may be responsible for electroweak symmetry breaking, as in
technicolor and topcolor models.  Examples of non-QCD-like theories
are the fixed-point models which have been discussed
recently.\cite{pmtechni} Third, it is important to test numerically
some of the analytic predictions of supersymmetric gauge theory, in
part to close some remaining loopholes in the analytic arguments.  In
addition, there are possible applications to condensed matter.

Four-dimensional supersymmetric gauge theories are good toy models for
non-supersymmetric QCD and its extensions.  Some of these theories
display confinement; of these, some break chiral symmetry but others
do not.\cite{nsexact,powerholo}  The mechanism of confinement
occasionally can be understood using a weakly-coupled ``dual
description''~\cite{nsewone,DS,NAD} (an alternate set of variables for
describing the same physics.)  However, what is more striking is that
most of these theories do not confine!~\cite{NAD}  Instead, their
infrared physics is governed by other, unfamiliar phases, often
described most easily using dual variables.  (Here and throughout,
``phase'' refers to the properties of the far infrared physics at zero
temperature.)  The phase diagram, as a function of the gauge group,
matter content, and interactions, is complex and intricate.\cite{NAD}

I will show that these theories have important connections with
non-supersymmetric theories.  While direct applications to QCD are
few, there are nonetheless important qualitative insights which can be
gained.  Explicitly broken \nfour\ and \ntwo\ supersymmetric
Yang-Mills (SYM) theories give windows into confinement.  Chiral
symmetry breaking can be explored in \none\ SYM and in
supersymmetric QCD (SQCD).  The large-$\nc$ limit offers the
connection between an ordinary field theory and a string theory on a
space of negative curvature,\cite{gubkleb,maldacon,GKP,ewAdS} and an
``orbifold conjecture'' relating a non-supersymmetric QCD-like theory to
\none\ SYM.  I'll also make some comments about avenues that
lattice QCD theorists might want to pursue.

\section{Pure \none\ Super-Yang-Mills}

Let us begin with a discussion of the pure \none\ \susic\ Yang-Mills
(SYM) theory --- simply $SU(N)$ gauge theory with a vector boson
(gluon) $A_\mu$ and a Majorana spinor (gluino) $\lam_\al$, and
Lagrangian
\bel{noneLgr}
\Lgr = {1\over 2g^2} \left[{\rm tr}\ F_{\mu\nu}F^{\mu\nu} 
+ i\bar\lambda D\slash \lambda \right]
\ee
Quantum mechanically, SYM bears some resemblance both to
ordinary non-supersymmetric YM and QCD, as it exhibits both
confinement and chiral symmetry breaking.

Let us examine a bit more closely the issue of chiral symmetry
breaking.  The classical $U(1)$ axial symmetry $\lam\rarr\lam
e^{i\sigma}$ is anomalous; since $\lam$ has $2N$ zero modes in the
presence of an instanton, only a $\ZZ_{2N}$, under which $\lambda
\rarr \lambda e^{i\pi n/N}$, is anomaly-free.  The dynamics causes a
gluino condensate $\vev{\lam\lam}$ to form, breaking this discrete
axial symmetry further to a $\ZZ_2$ under which $\lam\rightarrow
-\lam$.  This leads to $N$ equivalent vacua with
$\vev{\lam\lam}=\Lam^3 e^{i(2\pi n/N)}$.  Note that an instanton in
this theory comes with a factor $\Lam^{3N}$; the form of the gluino
condensate suggests it is induced by fractional instantons, indeed,
by $1/N$ of an instanton.

\begin{figure}
\centering
\epsfxsize=1.0in
\hspace*{0in}\vspace*{0in}
\epsffile{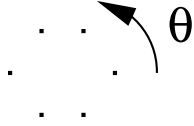}
\caption{The vacua of SYM theory are rotated
by the theta angle.}
\label{fig:vacua}
\end{figure}
\begin{figure}
\centering
\epsfxsize=2.3in
\hspace*{0in}\vspace*{0in}
\epsffile{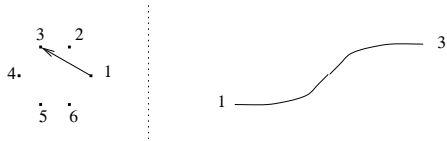}
\caption{Domain walls can interpolate between
vacua of SYM theory.}
\label{fig:walls}
\end{figure}

If we shift the theta angle by $\theta\rarr\theta+\alpha$, then
$\Lambda^3\rarr\Lambda^3 e^{i\alpha/N}$, and the $N$ equivalent vacua
(\FFig{vacua}) are rotated by an angle $\alpha/N$.  Any given vacuum
only comes back to itself under $\theta\rarr \theta + 2\pi N $, but
the physics is invariant under $\theta\rarr \theta + 2\pi $.  For
fixed $\theta$, the theory can have domain walls separating regions in
different vacua; as in \FFig{walls}, the condensate
$\vev{\lambda\lambda}$ can change continuously from $\Lambda^3e^{2\pi
i n/N}$ to $\Lambda^3e^{2\pi i n'/N}$.

Since \none\ SYM confines, it exhibits electric flux tubes.  As in
ordinary YM, the confining ``strings'' carry charge in the center of
the gauge group; for $SU(N)$ they carry quantum numbers $k$ in
$\ZZ_N$.  In later sections, I will discuss these flux tubes, their
charges, and their tensions $T_k$ in much greater detail.

It is important to note that similar properties --- equivalent
discrete vacua, domain walls, confining strings, and possible
fractional instantons --- should appear (and should be searched for)
in certain non-supersymmetric gauge theories with fermions.  In fact,
we expect them to occur in a set of models I will discuss in a later
section --- in so-called ``orbifolds'' of \none\ SYM theory.

What happens if we break \susy\ by adding a mass $m$ to the gaugino?

\bel{gauginomassLgr}
\Lgr \rarr \Lgr + m \lambda\lambda
\ee
The degeneracy of the $N$ SYM vacua will be broken; which one has
lowest energy will depend on the phase of $m$.  As
its phase is rotated, the preferred vacuum will shift from one to
the next. All domain walls become unstable (except for special choices
of arg$(m)$ for which the two lowest vacua become degenerate.)
However, the energy gap, the spectrum, and the general features of
confinement will not be greatly altered if $|m|\ll|\Lambda|$.

An issue for study using lattice methods is whether there are any
qualitative transitions in the physics of the theory as $m$ increases
and pure YM is recovered. The question is whether the properties of
SYM --- for example, the physics leading to confinement --- is
relevant for understanding pure YM.  Clearly, this rests in part on
whether the two theories are continuously connected, or whether they
are separated in parameter space by a phase transition.  For example,
consider the tensions $T_k$ of the flux tubes as a function of the
mass $m$; do they change continuously, or do they show a jump at some
value of $m$?  It would also be interesting to understand the behavior
of $\vev{\lam\lam}$ as a function of $m$.

Before returning to a study of confinement in SYM, I now want to turn
to supersymmetric gauge theories with matter, and some of their
remarkable properties.

\section{Gauge Theories with Matter}

Supersymmetric gauge theories teach us that the matter content of a theory
plays an essential role in determining its basic physics.  In
particular, the phase at zero temperature of a \gt\ (that is, its
properties in the far infrared) depends in a complicated way on (1) its
gauge group $G$, (2) its matter representations $R$, and (3) the
self-interactions $\Lgr_m$ of the matter, including non-renormalizable
interactions. Recent work has shown that the phase structure of \none\
\susic\ theories is complex and intricate, and it is reasonable to
expect that the same will be true of non-\susic\ theories.

Among the surprises discovered in the \susic\ context are that there
are new and unexpected phases unknown in nature or previously thought
to be quite rare.  It also appears that duality is a fundamental
property of field theory (and also of gravity, and even between
gravity and gauge theory!)  Certain accepted or at least popular lore
has been refuted as well: the beta function does {\it not}, by itself
anyway, determine the phase of a gauge theory; confinement does not
always cause chiral symmetry breaking; and fixed points in four
dimensions are not at all rare --- in fact, they are commonplace!

\subsection{Phases of SQCD}

Supersymmetric QCD consists of \none\ $SU(N)$ SYM with $\nf$
flavors of massless scalars (squarks) $Q$ and fermions (quarks) $\psi$ 
in the fundamental representation, along with anti-squarks and 
anti-quarks $\tilde Q$ and $\tilde \psi$ in the anti-fundamental
representation.  The Lagrangian is
$$
\Lgr = \Lgr_{SYM} + |DQ|^2 + i\bar\psi D\slash \psi + |D\tilde Q|^2 +
 i\overline{\tilde \psi} D\slash \tilde \psi +\ {\rm some\ interactions\ .}
$$

\begin{figure}
\centering
\epsfxsize=2.7in
\hspace*{0in}\vspace*{0in}
\epsffile{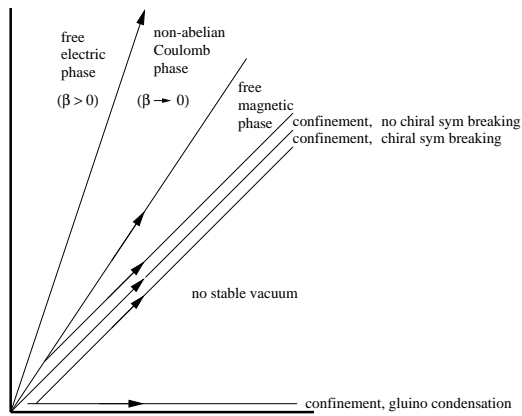}
\caption{The phases of SQCD, as a function of $N_f$ (vertical axis) and $N$ (horizontal axis.)}
\label{fig:phases}
\end{figure}

Following the work of Seiberg~\cite{NAD} and
others,\cite{kinstwo,kinsrev,kutsch,ils,gtwo} we now know that SQCD
has many phases.  \FFig{phases} shows the phase diagram of the
infrared behavior of the theory as a function of $N$ and $\nf$, as
determined by Seiberg.\cite{NAD} (Note that supersymmetric theories
generally have continuous sets of inequivalent vacua, called ``moduli
spaces;'' for each theory I have only listed the phase of the vacuum
with the largest unbroken global symmetry.  Also, unless otherwise
specified, I am always considering the case of strictly massless
matter.)

1) The Free Electric Phase:
When $\nf>3N$, the theory has a positive beta function and flows to
weak coupling.  The theory is free in the infrared in terms of
the original variables.

2) The Non-Abelian Coulomb Phase: When $3N>N_f>{3\over2}N$, the theory
is asymptotically free, but its beta function hits a zero at a finite
value of the gauge coupling.  The low-energy theory is an infrared
fixed point, an interacting conformal field theory, which has no
particle states.  Its operators have non-trivial anomalous dimensions,
some of which can be exactly computed. The theory has a ``dual
description'' using ``magnetic'' variables: in these variables, the
conformal theory is seen as an infrared fixed point of a {\it
different} gauge theory, one with $SU(N_f-N)$ gauge group and $N_f$
flavors as well as $N_f^2$ gauge-neutral massless fields.  

3) The Free Magnetic Phase: When ${3\over2}N\geq \nf\geq N+2$, the
theory flows to strong coupling in the infrared. The low-energy
physics can be described in terms of composite fields, of spin $0$,
$\half$, and $1$, many of which are non-polynomial --- indeed,
non-local --- in terms of the original fields.  The theory describing
these composites is again a gauge theory, with its own conceptually
separate gauge symmetry! Specifically, the theory of these ``dual''
fields is an $SU(\nf-N)$ gauge theory with $N_f$ flavors and $N_f^2$
neutral massless fields.  This dual theory has a positive beta
function, and so the magnetic degrees of freedom are infrared-free.
Thus, while the theory is strongly coupled in terms of
the original variables, these magnetic ``quasi-particles'' give a good
perturbative description of the low-energy physics.

4) Confinement Without Chiral Symmetry Breaking: For $N_f=N+1$, the
theory confines. In this case the low energy theory is an
infrared-free {\it linear} sigma model, describing composite mesons
and baryons (of spin $0$ and $\half$) that are polynomial in the
original fields.

5) Confinement With Chiral Symmetry Breaking: For $N_f=N$, the theory 
again confines. Again the light particles are massless mesons
and (scalar) baryons, but now the theory describing them breaks chiral
symmetry and becomes a non-linear sigma model.

6) $0<N_f<N$: When $\nf=N-1$, instantons generate an unstable potential for the
squarks; a similar effect, due to gaugino condensation (or perhaps
fractional instantons?) occurs for $N-2\geq \nf\geq 1$.   There is
no stable vacuum if the quarks and squarks are massless.

For $N_f=0$, we have pure \none\ SYM, which is in phase (5).

As another example, consider $SO(N)$ with $N_f$ flavors in the
vector representation, whose dual is
$SO(N_f-N+4)$.\cite{NAD,kinstwo}  In this case we have phase (1),
$N_f\geq 3(N-2)$; phase (2), $3(N-2)>N_f>{3\over2}(N-2)$; phase
(3), ${3\over2}(N-2)\geq N_f >N-3$; phase (4), $N_f=N-3, N-4$.
As before $N_f<N-4$ has no vacuum, except $N_f=0$ which has vacua
with phase (5).

I will now elaborate on these phases, and illustrate the dependence of
the phase structure on the details of the theory.

\subsection{Non-Abelian Coulomb Phase}

\begin{figure}
\centering
\epsfxsize=2.9in
\hspace*{0in}\vspace*{0in}
\epsffile{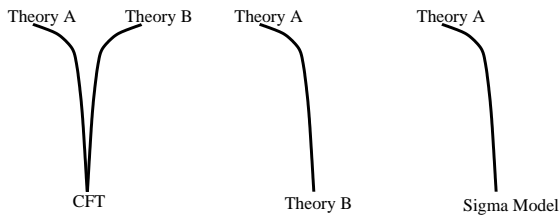}
\caption{(a) Non-abelian Coulomb Phase: two gauge theories flow to the
same conformal fixed point. (b) Free Magnetic Phase: gauge theory A
becomes strongly coupled, and at low-energy mysteriously turns into
gauge theory B.  (c) Confining Phase: theory A becomes strongly
coupled, and mysteriously turns into a theory of mesons and baryons.}
\label{fig:flow}
\end{figure}

This phase can be seen in perturbation theory, both in QCD and SQCD,
when the number of colors and flavors is large and the one-loop beta
function is extremely small by comparison; such fixed points are often
called Banks-Zaks fixed points,\cite{bz} though they were discussed by
earlier authors as well.  What is new here is that this phase exists
far beyond perturbation theory into an unexpectedly wide range of
theories.  Furthermore, these fixed points exhibit duality: there
exist multiple gauge theories which flow to the same conformal field
theory, as illustrated schematically in \FFig{flow}a.  Thus, there are
multiple sets of variables by which the conformal field theory may be
described.  This is analogous to Olive-Montonen duality in \nfour\
SYM, which is a conformal field theory; more on this later.

\subsection{Free Magnetic Phase}

This spectacular phase was entirely unexpected.  Here, the theory
flows to strong coupling, and its infrared physics is described by
weakly coupled composite matter and gauge fields.  These composites
are non-local with respect to the original degrees of freedom, and
have an unrelated gauge symmetry.  The duality transformation which
acts on the infrared fixed points of the non-abelian Coulomb phase operates in
the free magnetic phase by exchanging the fundamental theory of the
ultraviolet with the infrared-free composite theory, as illustrated in
\FFig{flow}b.  Note that the free magnetic phase is {\it not} a
confining phase.\cite{spinmono,ykis,nonBPS}.

\subsection{Confinement with and without chiral symmetry breaking}

In the confining theories, the low-energy description is a theory of gauge
singlets built from polynomials in the original degrees of freedom.
The main difference between chiral-symmetry-preserving and -breaking
theories is in the interaction Lagrangian, which determines the
symmetries preserved by the vacuum.

As in the free magnetic phase, the duality transformation exchanges
the ultraviolet theory with the infrared one --- the quarks and gluons
of the gauge theory are exchanged with the mesons and baryons of the
linear or non-linear sigma model (\FFig{flow}c).  At this point duality
begins to resemble what is done in real-world QCD when we rewrite the
theory in term of hadrons and the chiral Lagrangian.  This strongly
suggests that the QCD/chiral-Lagrangian ``duality'' transformation is
conceptually related to electromagnetic duality, its generalization to
Olive-Montonen duality, and its lower-dimensional cousins.  I will
make this more precise in later sections.

Note that the word ``confinement'' has been used loosely above, and
it is essential to make an important distinction. The
cases $SU(N)$ with $N_f=N+1,N$ are examples of
``complementarity'', where the confining and Higgs phases are actually
two regions in a single phase.\cite{compl}  There is no Wilson loop
with an area law; all sources can be screened by the massless fields,
and so no confining string can form.  By contrast, a spinor-valued
Wilson loop detects the confinement in $SO$ theories with vectors,
while Wilson loops in, for example, the ${\bf N}$ representation,
can detect the confinement in the pure $SU(N)$ and $SO(N)$ SYM.
In these cases the notion of confinement is precise.

\subsection{Dependence of Phase on Gauge Group and Matter Content}
While the phase structures of different gauge theories roughly resemble
one another, they are not by any means identical.  Consider, for
example, $SO(8)$ gauge theory with six fields in the vector
representation.  This theory is in the free magnetic phase, with an
abelian dual gauge group.\cite{kinstwo}  If we instead take $SO(8)$
with five vectors and one spinor --- this theory has the same one-loop
beta function as the previous one --- then the theory is confining,
and has a vacuum which does not break its chiral symmetries.  As
another example, compare $SU(N)$ SQCD with $\nf=N$ to \ntwo\ $SU(N)$
SYM.  The two theories have the same beta function, but the first is
in the confining phase with chiral symmetry breaking, while the second
is in the free magnetic phase.
\subsection{Dependence of Phase on Interactions}
The phase of a theory also depends strongly on the interactions
between the matter fields.  For example, consider $SU(4)$ SQCD with
$\nf$ flavors of squarks $Q_i$, anti-squarks $\tilde Q^i$, quarks and
anti-quarks $\psi_i,\tilde \psi^i$.  Consider taking a
``superpotential'' $W=yQ_1Q_2Q_3Q_4$, $y$ a coupling with
dimensions of inverse mass;  this adds the following
perturbation to the Lagrangian:
\bel{dimsix}
\left[
\sum_{i,j} {\partial^2 W\over \partial Q_i\partial Q_j}\psi_i\psi_j
+ h.c.
+\sum_i \left|{\partial W\over \partial Q_i}\right|^2
\right] \ .
\ee
Although this perturbation is irrelevant at weak coupling, it can be
important in the infrared, depending on $\nf$.  In particular,
\begin{itemize}
\item
For $\nf=8$, the theory is in the non-abelian coulomb phase: the
perturbation \eref{dimsix} is irrelevant, as in the classical limit.
\item
For $\nf=7$, the theory without the perturbation is in the non-abelian
coulomb phase, but the perturbation is relevant and drives the theory
to a different conformal field theory; thus the phase is unchanged but
the particular fixed point is different.  
\item
For $\nf=6$, the theory
without the perturbation is in the free magnetic phase, but the
perturbation drives the theory to an interacting fixed point and thus
into the non-abelian coulomb phase.  
\item
For $\nf=5$, the theory is
confining; the perturbation obviously breaks some chiral symmetries,
but even more are broken dynamically as a result of the perturbation.
\end{itemize}

\subsection{Concluding Comments}

Aside from the examples discussed here, many others are known.  For
most understood gauge theories, the phase diagrams --- a free electric phase
above, a conformal phase below, possibly a free magnetic phase and
usually confining phases, with or without chiral symmetry breaking ---
are qualitatively similar, although the details differ in very interesting
ways.  Various new phenomena have been uncovered along the way.
But most \none\ \susic\ field theories are not understood, and much
work remains to be done.

The most remarkable aspect of these phase diagrams is that they show
that the phase of a theory strongly depends on its gauge group $G$,  its
massless matter representations $R$, and 
its interactions $\Lgr_{int}$, renormalizable and non-renormalizable.
The dependence on $R$ goes far beyond the mere contribution of the
matter to the beta function; the matter fields are clearly more than
spectators to the gauge dynamics.  (A quenched approximation could not
reproduce this phase structure.)  The dependence on non-renormalizable
interactions is familiar from technicolor theories: a higher-dimension
operator, though irrelevant in the ultraviolet, may become relevant in
the infrared and control the physics of the low-energy theory.

It is also worth noting that while calculational techniques exist for
studying the infrared physics in most of these phases, the non-abelian
coulomb phase requires an understanding of four-dimensional
superconformal field theory.  Techniques in this subject are still
being explored~\cite{CFT} and there is much left to be learned.

Given this complexity is present in SQCD, why should it not be found
in non-\susic\ gauge theories?  The free electric phase, conformal or
non-abelian coulomb phase, and confining/chiral-symmetry-breaking
phase certainly arise.  It would be remarkable indeed if the ubiquity
of the conformal phase, the possibility of confinement without chiral
symmetry breaking, and the existence of a free magnetic phase could be
demonstrated.  There might also be as yet unknown phases that do not
occur in \susic\ theories.

 More specifically, we should seek to answer the following question:
what is the phase of QCD as a function of $G$, $R$ and $\Lgr_{int}$?
Unfortunately the answer cannot be learned from the supersymmetric
theories: the process of breaking supersymmetry leads to ambiguities
in the duality transformations.  We therefore need new tools, both
analytical and numerical.  This is clearly an area for lattice gauge
theory, but it is not easy to study the renormalization group flow
over large regions of energy using the lattice.  Additional analytic
work is needed to make this more tractable.  I hope some readers
will be motivated to consider this problem!

 It should be stressed that this is not merely an academic question.
It is possible that the correct theory of electroweak supersymmetry
breaking (or of fermion masses, etc.) has not yet been written down.
Perhaps a modified form of technicolor or something even more exotic
will appear in the detectors of the Large Hadron Collider, in a form
that we will be unable to understand unless the questions raised
above are addressed in the coming years.

\section{Unification of Dualities in \none\  Supersymmetry}

One of the most remarkable results of the recent developments
is that many strong-coupling phenomena previously thought to be
unrelated have turned out to be profoundly linked.  This 
is best understood by looking at a variety of ``duality transformations'',
non-local changes of variables which now seem to appear everywhere
we look for them.  Let us begin by listing a few: 

\

Electric-Magnetic (EM): this is the usual duality transformation of
the Maxwell equations without matter, which can be trivially extended
to \nfour, $2,1$ supersymmetric Maxwell theory.  The electric and
magnetic gauge groups are $U(1)_e$ and $U(1)_m$ (note these two
symmetry groups are {\it completely distinct} transformations on the
non-locally related electric and magnetic gauge potentials.)  The
electric and magnetic couplings are $e$ and $4\pi/e$.  (The last
relation is modified for non-zero theta angle.)

Dual Meissner (DM): for abelian gauge theory, or for a non-abelian
gauge theory which breaks down to an abelian subgroup.  Such a theory
often has magnetic monopoles, which are described as locally and
minimally coupled to a magnetic photon.  If the monopoles condense,
breaking the magnetic gauge symmetry, the magnetic photon obtains a
mass, screening magnetic flux and confining the electric flux of the
original description of the theory.

Olive-Montonen~\cite{omdual,gom,osborn} (OM): the EM case for \nfour\
\susy, extended to a non-abelian gauge group $G_e$.  The magnetic
variables also are an \nfour\ \susic\ gauge theory and have a gauge
group $G_m$.  The theory is conformal and has a quantum-mechanically
dimensionless (non-running) coupling constant $g$.  The coupling in
the magnetic description of the theory is $4\pi/g$.  (The last
relation is modified for non-zero theta angle.)

Generalized Dual Meissner~\cite{rdew,spinmono} (GDM): similar to the DM
case, but where both the electric and magnetic gauge groups $G_e$ and
$G_m$ are non-abelian.  Condensing magnetically charged monopoles
again break $G_m$, screen magnetic flux and confine electric flux.

Seiberg-Witten pure~\cite{nsewone} (SWp): for pure \ntwo\ \susic\
Yang-Mills theory.  The electric theory has gauge group $G$ of rank
$r$, whose maximal abelian subgroup is $[U(1)^r]_e$.  EM duality
applies to each $U(1)$; the magnetic theory has gauge group
$[U(1)^r]_m$.

Seiberg-Witten finite~\cite{nsewtwo} (SWf): for a finite \ntwo\ \susic\
gauge theory with matter.  Very similar to OM above, but in general
the relation between $G_e$ and $G_m$ differs from the OM case.

QCD and the Sigma Model (QCD$\sigma$): here a strongly-coupled,
confining QCD or \none\ SQCD theory is described in terms of gauge
singlets, using a linear or non-linear sigma model.  This is not
always considered a ``duality'', but as we will see, it should be.

The main point of this section is to emphasize that all of these
dualities are linked together~\cite{NAD,kinsrev} by results in \none\
\susy.  This can be easily seen using the duality of \none\ $SO(N)$
gauge theories with $N_f$ fields in the vector representation; as
mentioned earlier, such theories are dual to $SO(N_f-N+4)$ with
$N_f$ vectors and $N_f(N_f+1)/2$ gauge singlets.\cite{NAD,kinstwo}

Consider first $SO(2)$ without matter ($N_f=0$).  The dual theory is
again $SO(2)$ without matter --- EM duality --- which justifies
referring to the dual theory as ``magnetic''.

Next, consider $SO(3)$ with one triplet; its magnetic dual is $SO(2)$
with fields of charge $1,0,-1$ coupled together.  These theories are
both \ntwo\ \susic; the electric theory is the pure \ntwo\ $SU(2)$
theory studied by Seiberg and Witten, and the magnetic dual is the
theory of the light monopole which serves as its low-energy
description.\cite{nsewone} Since a mass for the triplet leads to
confinement of $SO(3)$ via abelian monopole
condensation,\cite{nsewone} this example gives both SWp and DM
duality.

Finally, take $SO(3)$ with three triplets, whose dual is $SO(4)\approx
SU(2)\times SU(2)$, with three fields in the ${\bf 4}$ representation
and six singlets.  If the triplets are massive, the quartets of the
dual theory condense, $SO(4)$ is broken, and confinement of $SO(3)$
occurs~\cite{NAD} --- GDM duality.  The low-energy description below
the confinement scale is given by this broken $SO(4)$
theory~\cite{NAD}, a non-linear sigma model --- QCD$\sigma$ duality.
And if instead the triplets are massless and are given the
renormalizable interactions which make the theory \nfour\ \susic, the
dual theory is consistent with OM duality:~\cite{omdual,nsewtwo} one of
the $SU(2)$ subgroups of $SO(4)$ confines, leaving a single $SU(2)$
factor with three triplets coupled by the required \nfour\ \susic\
interactions.\cite{kinstwo,kinsrev}

\begin{figure}
\centering
\epsfxsize=3in
\hspace*{0in}
\epsffile{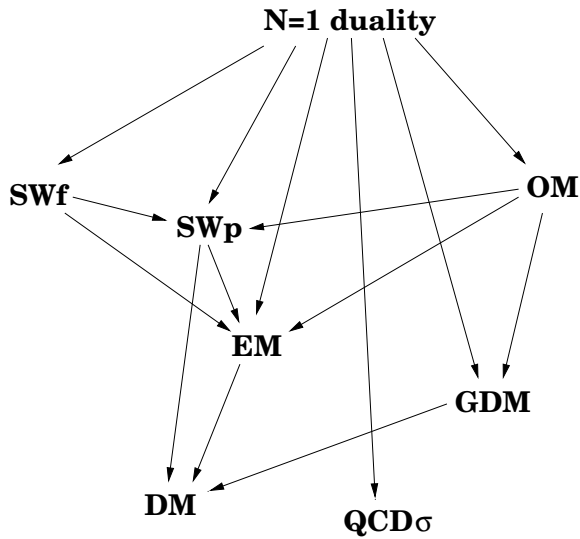}
\caption{Interrelation between types of duality.}
\label{fig:dualities}
\end{figure}

Thus, \none\ duality links all of the four-dimensional dualities on
the list together, showing they are manifestations of a single
phenomenon.  Figure \ref{fig:dualities} shows a cartoon of the
connections between these theories.  I have omitted, both from my list
and from this figure, other types of field theory duality that were
not known until recently.  Also not shown are the relations between
these dualities and those of the-theory-formerly-known-as-string
theory, sometimes termed M theory.  The situation brings to mind the
old story of the blind men and the elephant, in which each man feels
one part of the elephant --- the ear, or the trunk, or the tail ---
and erroneously assigns it a special significance, not realizing that
the apparently different parts are connected to a larger whole.
Having discovered the existence of a larger whole, we find ourselves
compelled to explain duality in a unified way.  There
have been some recent developments
in this subject.\cite{superQED,HoriVafa}

I want to emphasize to those who are not familiar with this subject
that this picture, while not proven, is by no means speculative.  The
circumstantial evidence in its favor --- a vast number of consistency
conditions --- is completely overwhelming; one could easily give
twenty lectures on this subject.  A proof is still badly needed,
however; the underlying meaning of duality remains mysterious, and no
field theoretic formulation is known which would make it self-evident
(except in the EM case, of course.)

\section{Olive-Montonen Duality, Confinement, and the QCD String}

What drives confinement?  In particular, why is \none\ SYM confining?
This question cannot be addressed directly in SYM, because this theory
does not have a weakly-coupled dual description. However, there is a
trick for studying this question --- within limits, as discussed
further below.  The trick is the following.  One can add additional
massive matter to SYM without leaving its universality class (note
supersymmetry and holomorphy ensure this.)~\cite{powerholo} In
particular I will study broken \ntwo\ SYM~\cite{nsewone,DS} and broken
\nfour\ SYM~\cite{cvew,rdew,ykis} gauge theories, which have the same
massless fields as \none\ SYM.  These theories have duality
transformations which allow their dynamics to be studied using
weakly-coupled magnetic descriptions.  In these descriptions there are
monopoles in the theory, which condense, thereby causing
confinement.\cite{nsewone} The confining flux tubes appear as
string solitons.  The picture which emerges appears to support the old
lore that the Dual Meissner effect is responsible for confinement.

Before beginning the pedagogical presentation of these arguments, let
me summarize the results to be obtained below.  The \ntwo\ $SU(N)$ SYM
theory, with strong coupling scale $\Lam$, has a dual description as
an abelian $U(1)^{N-1}$ \gt;~\cite{nsewone,sun} the monopoles of the
$SU(N)$ description are electrically charged under the dual
description.  When masses $\mu\ll\Lam$ are added to the extra matter,
so that the only massless fields are those of \none\ SYM, the
monopoles develop expectation values.  The dual description of this
condensation involves nothing more than $N-1$ copies of the Abelian
Higgs model, and so gives $N-1$ solitonic Nielsen-Olesen
strings,\cite{nsewone,DS} each carrying an integer charge.  Because of
this monopole condensate, electrically charged sources of $SU(N)$ are
confined, and the flux tubes which confine them are the solitonic
strings of the dual description.  However, the confining strings are
problematic.\cite{DS,hsz} Although they carry an exact $\ZZ_N$
symmetry, they also each carry an (approximate) integer charge,
violated only at the scale $\Lam$ which is large compared to the
string tension.  This extra symmetry leads the theory to exhibit not
one but $N/2$ metastable Regge trajectories --- and thus the theory
does not have the same properties as YM or \none\ SYM.  As
$\mu\rarr\Lam$, the extra symmetry begins to disappear, but at the
same time the magnetic description becomes strongly coupled, so no
reliable dual description can be given in the regime where the theory
behaves as \none\ SYM is expected to do.

\begin{figure}
\centering
\epsfxsize=2.5in
\hspace*{0in}\vspace*{0in}
\epsffile{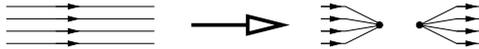}
\caption{The annihilation of $N$ flux tubes using baryon vertices;
this transition would be inhibited if a good dual description of
YM theory involved $U(1)^{N-1}$.}
\label{fig:noannih}
\end{figure}

Note these properties are characteristic of abelian projection
approaches to confinement.  If we project $SU(N)\rarr U(1)^{N-1}$,
dynamically or otherwise, then abelian monopole condensation leads to
Nielsen-Olesen strings, each with its own approximately conserved
integer charge.  This unavoidable charge inhibits annihilation of $N$
identical strings (as in \FFig{noannih}) which both leads to an
overabundance of metastable hadrons~\footnote{Here and elsewhere, by
``hadrons of Yang-Mills theory'' I mean mesons and baryons built from
quarks whose masses are large enough compared to $\Lambda$ that they
do not greatly affect the infrared dynamics.}  and to difficulty in
forming baryons.\cite{hsz} This is a serious problem for abelian
projection approaches to YM and QCD.

The situation in broken \nfour\ SYM~\cite{ykis} is much more
satisfactory.  \nfour\ $SU(N)$ SYM is a conformal field theory, with
gauge coupling $g$.  Its magnetic description, also an \nfour\
conformal field theory, has gauge group $SU(N)/\ZN$ and coupling
$1/g$; thus, if $g\gg 1$, the magnetic description is weakly coupled.
Adding masses $\mu$ to all but the fields of \none\ SYM causes the
scalars of the magnetic description to condense, breaking the dual
gauge group completely.\cite{cvew,rdew}  This {\it non}-Abelian Higgs
model has string solitons, but because the fundamental group of
$SU(N)/\ZN$ is $\ZN$, these strings carry a $\ZN$ charge,\cite{ykis}
in contrast to the integer charges found in the case of broken \ntwo\
SYM. In short, the electric description of the theory involves
confinement by electric flux tubes carrying $\ZN$ electric flux,
leading to a single Regge trajectory, in agreement with expectations
for \none\ SYM.  Notice that the associated dual description does not
resemble abelian projection: it is essential for the $\ZN$ charges of
the strings that the dual theory is {\it non-abelian}.  Since one
cannot obtain the dual $SU(N)/\ZN$ theory by a projection on the
$SU(N)$ theory --- the relation between the two is fundamentally
quantum mechanical --- it seems to me that abelian projection is
conceptually disfavored.

Note however that one cannot carry this logic all the way to the \none\ SYM
theory itself. To do so requires taking $\mu\rarr\infty,g\rarr 0$, but
in this limit the magnetic theory becomes strongly coupled and the
semiclassical discussion of the previous paragraph becomes unreliable.

Thus, a strong word of caution is in order here.  In particular,
although both broken \ntwo\ and broken \nfour\ are in the same
universality class as \none\ SYM, they are not equivalent to it.
While confinement and an energy gap are universal properties of all of
these theories, the monopoles which lead to confinement are {\it not}
universal.  The properties of the monopoles depend on the matter
that is added to SYM.  Again, this poses problems for abelian
projection approaches to confinement, or indeed for any semiclassical
description of the phenomenon.

\subsection{Electric Sources and Fluxes}

I begin with a review of basic and well-known facts about gauge
theories.  Consider a pure gauge theory with gauge group $G$.  Suppose
we have a source --- an infinitely massive, static, electrically
charged particle --- in a representation $R$ of $G$.  If we surround
the source with a large sphere, what characterizes the flux passing
through the sphere?  If $G$ is $U(1)$, the flux measures the electric
charge directly. However, in non-abelian gauge theories the gauge
bosons carry charge.  Since there may be a number (varying over time)
of gauge bosons inside the sphere, the representation under which the
charged objects in the sphere transform is not an invariant.  But, by
definition, the gauge bosons are neutral under the discrete group
$C_G$, the center of $G$.  It follows that the charge of $R$ under the
center {\it is} a conserved quantity, and that the total flux exiting
the sphere carries a conserved quantum number under $C_G$.

For example, in $SU(N)$ the center is a $\ZN$ group with the lovely
name of ``N-ality''.  A tensor $T^{a_1a_2\cdots a_p}_{b_1b_2\cdots
b_q}$ has N-ality $p-q$ mod $N$; in particular, the $k$-index
antisymmetric tensors carry N-ality $k$.

\ures Electric sources and fluxes in pure gauge theories carry a
conserved $C_G$ quantum number.  If the gauge group confines, then the
confining electric flux tubes will also carry this quantum number.

If the theory also contains light matter charged under $C_G$ but
neutral under a subgroup $C_m$ of $C_G$, then the above statements are
still true with $C_G$ replaced with $C_m$.  For example, if we take
$SU(N)$ with massless fields in the ${\bf N}$ representation, then
$C_m$ is just the identity, reflecting the fact that all sources can
be screened and all flux tubes break.  If we take $SO(10)$ with fields
in the ${\bf 10}$, then the center $\ZZ_4$ is replaced with
spinor-number $\ZZ_2$.  Sources in the ${\bf 10}$ will be screened and
have no flux tube between them, while sources in the ${\bf 16}$ or
${\bf \overline{16}}$ will be confined by a single type of flux tube.

\subsection{Confinement and Flux Tubes}

Like pure YM, \none\ SYM is confining; how does this discussion apply
to it? Since the gauge bosons and gauginos are all in the adjoint
representation, the electric flux tubes do carry charge in the center
of the gauge group; for $SU(N)$ they carry quantum numbers $k$ in
$\ZZ_N$.  What properties of these strings could we hope to predict?
The tension $T_k$ of an infinitely long confining string is a function
of $k,N,\Lam$, and we might hope to say something about it.  First,
note $T_k=T_{N-k},T_N=0$ by $\ZZ_N$ symmetry.  An electric source of
charge $k$ ({\it e.g.} one in a $k$-index antisymmetric tensor
representation) will be confined by a $k$-string (a string carrying
$k$ units of flux) or by a set of strings whose charges add to $k$ mod $N$.  The ratio $T_k/T_{k'}$ is a basic property of YM
and SYM, as fundamental as the glueball spectrum and easier for
theorists to estimate.  
I will discuss some predictions for these ratios later.

\subsection{Magnetic Sources and Fluxes}

Now let consider what happens to magnetic sources and fluxes
in gauge theories.  First, let's review some basic topology.
The $p$-th homotopy group of a manifold $\MM$, $\pi_k(\MM)$, is the
group of maps from the $p$-sphere into $\MM$, where we identify maps as
equivalent if they are homotopic in $\MM$.  All we will need for present
purposes are the following examples.  Suppose a Lie group $G$ has rank
$r$, so that its maximal abelian subgroup is $U(1)^r$; then
\bel{pitwo}
\pi_2[G] = {\bf 1}\ \  \Rightarrow \ \
\pi_2\left[G/U(1)^r\right] 
= \pi_1[U(1)^r] =  \ZZ\times \ZZ\times \cdots \times \ZZ
\equiv [\ZZ]^r  \ .
\ee
Similarly,
\bel{pione} 
\pi_1[G] = {\bf 1} \ \  \Rightarrow \ \ \pi_1\left[G/C_G\right] =
\pi_0[C_G] = C_G\ .  
\ee

We will need to investigate both monopole solitons and string solitons
below.  The classic monopole soliton is that of 't Hooft and of
Polyakov, which arises in $SU(2)$ broken to $U(1)$; in this case the
important topological relation is $\pi_2[SU(2)/U(1)] =
\pi_1[U(1)]=\ZZ$.  This leads to a set of monopole solutions carrying
integer charge.  Note that the stability of, for example, the
charge-two monopole solution against decay to charge-one monopoles is
determined not by topology but by dynamics.  The situation is similar
for the Nielsen-Olesen magnetic flux tube of the abelian Higgs model;
here the relevant topological relation is $\pi_1[U(1)]=\ZZ$.  This
again leads to solutions with an integer charge, whose stability
against decay to minimally charged vortices is determined 
dynamically.

More generally, if we have a {\it simply connected} gauge group $G_0$
which breaks to a group $G$ at a scale $v$, there will be monopoles
carrying a quantum number in $\pi_2[G_0/G]$, of mass [radius]
proportional to $v$ [$1/v$].  Now imagine that we take $v\rarr\infty$.
In this limit the gauge group $G_0$ disappears from the system.  The
monopoles become pointlike and infinitely massive; their only
non-pointlike feature is their Dirac string, which carries a quantum
number in $\pi_1[G]$. In short, the solitonic monopoles become
fundamental Dirac monopoles in this limit. Note that since
$\pi_2[G_0/G]=\pi_1[G]$, the charges carried by the solitonic
monopoles and their Dirac monopole remnants are the same.  Since the
Dirac monopoles are heavy, we may use them as magnetic sources.

Let's further suppose that the gauge group $G$ is broken completely at
some scale $v'$.  In this case no Dirac strings can exist in the
low-energy theory, and so the monopoles allowed previously have
seemingly vanished.  However, solitonic magnetic flux tubes, carrying
charges under $\pi_1[G]$, will be generated; they will have tension
[radius] of order $v'^2$ [$1/v'$].  Their $\pi_1[G]$ quantum numbers
are precisely the ones they need to confine the $\pi_1[G]$-charged
Dirac monopole sources of the high-energy theory.  Thus, when $G$ is
completely broken, the Dirac monopoles disappear because they are
confined by flux tubes.

\ures Magnetic sources and fluxes in pure gauge theories carry a
conserved $\pi_1[G]$ quantum number.  If the gauge group is completely
broken, then the confining magnetic flux tubes will also carry this
quantum number.

\subsection{\nfour\ Supersymmetric Gauge Theory}

The next ingredient in this stew is \nfour\ \susic\ gauge theory,
consisting of one gauge field, four Majorana fermions, and six real
scalars, all in the adjoint representation.  It is useful to combine
these using the language of \none\ \susy, in which case we have one
vector multiplet (the gauge boson $A_\mu$ and one Majorana fermion
$\lambda$) and three chiral multiplets (each with a Weyl fermion
$\psi^s$ and a complex scalar $\Phi^s$, $s=1,2,3$.)

These fields have the usual gauged kinetic terms, along with
additional interactions between the scalars and fermions.  The
scalars, in particular, have potential energy
\be
V(\Phi^s) = \sum_{a=1}^{dim \ G} |D_a^2| + \sum_{s=1}^3 |F_s|^2
\ee
 where
\be
D_a = \left(\sum_{s=1}^3 [\Phi^{s\dag},\Phi^s]\right)_a
\ee
(here $a$ is an index in the adjoint of $G$) and
\bel{Fnomass}
F_s = \epsilon_{stu}  [\Phi^t,\Phi^u] \ .
\ee
Supersymmetry requires that $\vev{V(\Phi^s)}=0$, and so all $D_a$ and
$F_s$ must vanish separately.  The solution to these requirements is
that a single linear combination $\hat \Phi$ of the $\Phi^s$ may have
non-vanishing expectation value, with the orthogonal linear
combinations vanishing.  By global rotations on the index $s$ we may
set $\hat \Phi=\Phi^3.$ By gauge rotations we may make $\Phi^3$ lie in
the Cartan subalgebra of the group; we may represent it as a diagonal
matrix
\bel{vacs}
\vev{\Phi^3} = {\rm diag}(v_1,v_2,\cdots)
\ee
which (if the $v_i$ are all distinct) breaks $G$ to $U(1)^r$.
Since $\pi_2[G/U(1)^r]= [\ZZ]^r$ [see \Eref{pitwo}] the theory
has monopoles carrying $r$ integer charges under $U(1)^r$.
Quantum mechanically, the theory has both monopoles and dyons,
carrying $r$ electric and $r$ magnetic charges $(n_e,n_m)$.    

The space of vacua written in \Eref{vacs} is not altered by quantum
mechanics.  In the generic $U(1)^r$ vacuum, each $U(1)$ has EM
duality.  In a vacuum where some of the $v_i$ are equal, the gauge
group is broken to a non-abelian subgroup $\hat G$ times a product of
$U(1)$ factors; the low-energy limit of the non-abelian part is an
interacting conformal field theory.  The $U(1)$ factors have EM
duality, while the $\hat G$ factor has its non-abelian generalization,
OM duality.\cite{omdual,gom,osborn,nsewtwo}

\subsection{Olive-Montonen Duality}

The \nfour\ theory has a set of alternate descriptions, generated
by changes of variables (whose explicit form remains a mystery)
of the form
\be
{\bf T} \ : \ \tau\rarr\tau+1 \ (\theta\rarr\theta+2\pi) \ ; \
n_e\rarr n_e+n_m,\ n_m\rarr n_m \ ; \ G\rarr G \ ;
\ee
and
\bel{Stransf}
{\bf S} \ : \ \tau\rarr-{1\over\tau} \ 
 (g\rarr {4\pi\over g} {\rm \ if \ } \theta=0) \ ; \
n_e\leftrightarrow n_m \ ; \ G\rarr \tilde G \ .
\ee
Together ${\bf S}$ and ${\bf T}$ generate the group $SL(2,\ZZ)$, which
takes $\tau\rarr (a\tau+b)/(c\tau+d)$ for integers $a,b,c,d$
satisfying $ad-bc=1$.  Note that ${\bf T}$ is nothing but a rotation
of the theta angle by $2\pi$; it does not change the gauge group or
the definition of electrically charged particles, shifting only the
electric charges of magnetically charged particles.\cite{witeff} By
contrast, ${\bf S}$ exchanges electric and magnetic charge, weak and
strong coupling (if $\theta=0$),\cite{omdual} and changes the gauge
group~\cite{gom,osborn} from $G$ to its dual group $\tilde G,$ as
defined below.

The group $G$ has a root lattice $\Lambda_G$ which lies in an $r=$
rank$(G)$ dimensional vector space.  This lattice has a corresponding
dual lattice $(\Lambda_G)^*$.  It is a theorem that there exists a Lie
group whose root lattice $\Lambda_{\tilde G}$ equals
$(\Lambda_G)^*$.\cite{gom} Here are some examples:
\be 
\matrix{{SU(N)}\leftrightarrow SU(N)/\ZN \ ; & \ 
{SO(2N+1)}\leftrightarrow USp(2N) \ ; \cr 
{SO(2N)}\leftrightarrow SO(2N)   \ ; & \ 
{Spin(2N)} \leftrightarrow  SO(2N)/\ZZ_2 \ .}
\ee
Notice that this set of relationships depends on the global
structure of the group, not just its Lie algebra; $SO(3)$ (which
does not have spin-$1/2$ representations) is dual to $USp(2)\approx SU(2)$
(which does have spin-$1/2$ representations.)  These details are essential
in that they affect the topology of the group, on which OM duality depends.

In particular, there are two topological relations which are of great
importance to OM duality.  The first is relevant in the generic
vacuum, in which $G$ is broken to $U(1)^r$.  The electric charges
under $U(1)^r$ of the massive electrically charged particles (spin
$0,\half,1$) lie on the lattice $\Lambda_G$. The massive magnetic
monopoles ({\it also} of spin $0,\half,1$) have magnetic charges under
$U(1)^r$ which lie on the dual lattice
$(\Lambda_G)^*$.\cite{gom,osborn} Clearly, for the ${\bf S}$ transformation,
which exchanges the electrically and magnetically charged fields and
the groups $G$ and $\tilde G$, to be consistent, it is essential that
$\Lambda_{\tilde G} = (\Lambda_G)^*$ --- which, fortunately, is true.

The second topological relation is the one we will use below.  We have
seen that the allowed electric and magnetic sources for a gauge theory
with adjoint matter (such as \nfour) are characterized by quantum
numbers in $C_G$ and $\pi_1(G)$ respectively.  Consistency of the
${\bf S}$ transformation would not be possible were these two groups
not exchanged under its action.  Fortunately, it is a theorem of group
theory that~\cite{gom}
\bel{piGCG}
\pi_1(G) = C_{\tilde G} \ ; \ \pi_1(\tilde G)= C_G \ .
\ee
For example, $\pi_1[SU(N)] = C_{SU(N)/\ZN} = {\bf 1}$ while
$C_{SU(N)} = \pi_1[SU(N)/\ZN] = \ZN$.

\ures As a consequence of \Eref{piGCG} and the results of sections 5.1
and 5.2, the allowed magnetic sources of $G$ are the same as the
allowed electric sources for $\tilde G$, and vice versa.

\subsection{Breaking \nfour\ to \none}

 Now, we want to break \nfour\ \susy\ to \none.  Pure \none\ SYM,
like pure non-\susic\ YM, is a confining theory, and should contain
confining flux tubes.  The addition of massive matter in the adjoint
representation does not change this; heavy particles would only obstruct
confinement by breaking flux tubes, which adjoint matter cannot do.
We therefore expect that broken \nfour\ gauge theory, which is \none\
SYM plus three massive chiral fields in the adjoint representation,
should be in the same universality class as pure SYM: both should
confine, and both should have flux tubes carrying a $C_G$ quantum
number, as discussed in section 5.1.

We may break the \nfour\ symmetry by adding masses $m_s$ for
the fields $\Phi^s$; the $F_s$ functions of \eref{Fnomass} become
\bel{Fwmass}
F_s = \epsilon_{stu} [\Phi^t,\Phi^u] + m_s\Phi^s \ ,
\ee
so that $F_s=0$ implies $\epsilon_{stu}[\Phi^t,\Phi^u]=-m_s\Phi^s$. Up
to normalization these are the commutation relations for an $SU(2)$
algebra,\cite{cvew} which I will call $SU(2)_{aux}$.  If we take
$m_1=m_2=m$ and $m_3=\mu$ we obtain
\bel{Phivevs}
\Phi^1 = -i\sqrt{\mu m} J_x \ ;
\Phi^2 = -i\sqrt{\mu m} J_y \ ;
\Phi^3 = -i m J_z \ ,
\ee
where $J_x,J_y,J_z$ are matrices satisfying $[J_x,J_y]=iJ_z$, {\it
etc.}  Each possible choice for the $J$'s gives a separate, isolated
vacuum.\cite{cvew}

How does this work, explicitly, in $SU(N)$?  We can write the $\Phi^s$
as $N\times N$ traceless matrices, so the $J_i$ should be an
$N$-dimensional (generally reducible and possibly trivial)
representation of $SU(2)_{aux}$.\cite{cvew,rdew}  The trivial choice
corresponds to $J_i=0$; clearly if $\Phi^s=0$ the $SU(2)_{aux}$
commutation relations are satisfied.  We will call the corresponding
vacuum the ``unbroken'' vacuum, since the $SU(N)$ gauge group is
preserved.  Another natural choice is to take the $J_i$ in the
irreducible spin-${N-1\over2}$ representation of $SU(2)_{aux}$.  In
this case $SU(N)$ is completely broken; we will call this the ``Higgs
vacuum''.  We may also choose the $J_i$ in a reducible representation
such as
\be
J_i= \left[\matrix{\sigma_i & | &0\cr  --&-|-&--\cr 0& | &0 }\right] \ ;
\ee
here the $\sigma_i$ are the Pauli matrices.
In this case $SU(N)$ is partly broken.  There are many vacua like this
last one, but they will play no role in the physics below; we will
only need the unbroken vacuum and the Higgs vacuum.

\ures The classical analysis of this \none\ \susic\ $SU(N)$ gauge
theory with massive adjoint fields shows that it has isolated \susic\
vacua scattered about, with the unbroken (U) vacuum at the origin of
field space and the Higgs vacuum (H) at large $\Phi^s$ expectation
values [of order $m,\sqrt{m\mu}$, see \Eref{Phivevs}.]~\cite{cvew,rdew}

\subsection{OM Duality and the Yang-Mills String}

The above picture is modified by quantum mechanics.  In each vacuum,
strong dynamics causes confinement to occur in the unbroken
non-abelian subgroup, modifying the low-energy dynamics and generally
increasing the number of discrete vacua.  In the H vacuum, the gauge
group is completely broken and no non-trivial low-energy dynamics
takes place; it remains a single vacuum.  The U vacuum, by contrast,
splits into $N$ vacua --- the well-known $N$ vacua of
SYM~\cite{Nvacua} --- which I will call ${\rm D}_0,{\rm
D}_1,\cdots,{\rm D}_{N-1}$. In the ${\rm D}_k$ vacuum,
confinement occurs by condensation of dyons of magnetic charge 1 and
electric charge $k$.\cite{nsewone,rdew,DS,sun}  Since these vacua are
related~\cite{Nvacua} by rotations of $\theta$ by multiples of $2\pi$,
I will focus on just one of them.  It is convenient to study the ${\rm
D}_0$ vacuum (which I now rename the M vacuum) in which electric
charge is confined by magnetic monopole condensation.

Now, what is the action of OM duality on this arrangement?  The vacua
are physical states, and cannot be altered by a mere change of
variables; however, the {\it description} of each vacuum will change.
Specifically, when $\theta=0$, the ${\bf S}$ transformation, which
inverts the coupling constant and exchanges electric and magnetic
charge, exchanges the H vacuum of $SU(N)$ for the M vacuum of
$SU(N)/\ZN$ and vice versa.\cite{rdew}  To say it another way,
the confining M vacuum of $SU(N)$ can be equally described as the H
vacuum of $SU(N)/\ZN$, in which the monopoles of the $SU(N)$
description break the dual $SU(N)/\ZN$ gauge group.  This is the
Generalized Dual Meissner effect, in which both the electric and
magnetic gauge groups are non-abelian.

\ures OM duality exchanges the H and M vacua of $SU(N)$ broken \nfour\
with the M and H vacua of $SU(N)/\ZN$ broken \nfour.  Confinement in
the M vacuum of $SU(N)$ is described as the breaking of the
$SU(N)/\ZN$ gauge group in $SU(N)/\ZN$.\cite{rdew}

The existence of the Yang-Mills string now follows directly from
topology.  As we discussed in section 5.2, the complete breaking of a
group $G$ leads to solitonic strings carrying magnetic flux with a
quantum number in $\pi_1(G)$.  In this case, the breaking of
$SU(N)/\ZN$ in its H vacuum (the confining M vacuum of $SU(N)$) gives
rise to  strings with a $\ZN$ quantum number.  But magnetic flux tubes
of $SU(N)/\ZN$ are, by OM duality, electric flux tubes of $SU(N)$ ---
and so the confining strings of the $SU(N)$ theory's M vacuum, the
confining theory which is in the same universality class as $SU(N)$
SYM, carry a $\ZN$ quantum number.  This is in accord with the
considerations of section 5.1.  The relation \Eref{piGCG} is
responsible for this agreement of the $\ZN$ charges, and presumably
assures a similar agreement for all groups.

\ures OM duality gives a new picture for confinement in $SU(N)$ SYM:
it occurs via {\it non-abelian} dual monopole condensation, and leads
to confining strings with a $\ZN$ quantum number.

A cautionary remark is in order.  The description of confinement via
dual monopole condensation is not fully reliable, as it is only
appropriate if the $SU(N)/\ZN$ theory is weakly coupled.  In fact, we
want the $SU(N)$ theory to be weakly coupled in the ultraviolet, in
analogy with QCD.  The ${\bf S}$ transformation implies that the
$SU(N)/\ZN$ description should be {\it strongly} coupled in the
ultraviolet.  However, the existence of a soliton carrying a stable
topological charge is more reliable, especially since there are no
other objects carrying that charge into which these string solitons
can decay.  Having constructed the solitonic strings semiclassically
in some regime, we expect that they survive into other regimes in
which semiclassical analysis would fail. (A gap in the argument: could
the strings grow large and have zero string tension in the SYM
limit?)  In short, while the condensing monopole description
appropriate to broken \nfour\ SYM may not be valid for pure \none\
SYM, it does demonstrate the presence of confining strings in the
latter.  Whether anything quantitative can be said about these strings
is another matter, to be addressed below.

Should we expect this picture to survive to the non-\susic\ case?
Take the theory with \nfour\ \susy\ broken to \none, and futher break
\none\ \susy\ by adding an $SU(N)$ gaugino mass $m_\lambda\ll m,\mu$.
We cannot be sure of the effect on the dual $SU(N)/\ZN$ theory;
duality does not tell us enough. However, we know that the theory has
a gap, so this \susy-breaking can only change some properties of the
massive fields, without altering the fact that $SU(N)/\ZN$ is
completely broken.  The strings, whose existence depends only on this
breaking, thus survive for small $m_\lambda$.  To reach pure YM
requires taking $m,\mu,m_\lambda$ all to infinity.  It seems probable,
given what we know of YM physics, that the strings undergo no
transition as these masses are varied.  In particular, there is
unlikely to be any phase transition for the strings between pure SYM
and pure YM; this conjecture can and should be tested on the lattice.

\ures If the strings of SYM and of YM are continously related,
without a transition as a function of the gaugino mass, then the
arguments given above for SYM extend to YM, establishing a direct
link between OM duality of \nfour\ gauge theory and the confining
$\ZN$-strings of pure YM.

\subsection{Confinement According to Seiberg and Witten}

How does this picture of confinement differ from that of Seiberg and
Witten?  Where and why might it be preferable?

Seiberg and Witten studied pure \ntwo\ \susic\ $SU(2)$ gauge
theory.\cite{nsewone} They showed that the infrared quantum mechanical
theory could be understood as a $U(1)$ theory coupled to a magnetic
monopole, and that, when \ntwo\ \susy\ is broken to \none, the
monopole condenses, confining the $SU(2)$ degrees of freedom.  The
picture generalizes~\cite{sun} to $SU(N)$, where the infrared physics involves
$U(1)^{N-1}$ coupled to $N-1$ monopoles, whose condensation drives
confinement.  This was studied in detail by Douglas and Shenker.\cite{DS}

This physics is contained in a particular regime of the broken \nfour\
\susy\ gauge theory discussed above.  If we take $\mu=0$, then the
theory is \ntwo\ \susic, and, as seen from \Eref{Phivevs}, the H
vacuum of $SU(N)/\ZN$ has its gauge group broken only to $U(1)^{N-1}$.
Now take $m$ exponentially large and the coupling at the scale $m$
small, so that the strong coupling scale $\Lambda$ of the low-energy
theory is finite, and let $\mu$ be non-zero but small compared to
$\Lambda$. The low-energy confining vacua of the $SU(N)$ theory will
then be the vacua studied by Douglas and Shenker.  The magnetic
description of the theory (using OM duality) will have the gauge group
$SU(N)/\ZN$ broken at a high scale to $U(1)^{N-1}$, which in turn is
broken completely at a low scale; see \Eref{Phivevs}.  The second step
in this breaking leads to confinement of $SU(N)$ fields.  If we take
$m$ to infinity, then the $SU(N)/\ZN$ gauge group disappears from the
theory.  The magnetic theory is merely $U(1)^{N-1}$ broken to nothing,
as in Douglas and Shenker.\cite{sun,DS} (Note that I am cheating a bit
here, as the abelian theory requires a cutoff; I'll fix this below.)

Since the magnetic theory is $U(1)^{N-1}$, dual to the maximal abelian
subgroup of $SU(N)$, the pure \ntwo\ theory exhibits a dynamical form
of abelian projection.\cite{DS} The monopoles, whose condensation
drives confinement when \ntwo\ \susy\ is weakly broken, are purely
abelian.  Given the long-standing arguments using the monopoles of
abelian projection as a mechanism for explaining confinement, why
should this disturb us?

The problem lies with the quantum numbers of the strings.  The
$U(1)^{N-1}$ magnetic theory consists of $N-1$ copies of the abelian
Higgs model, each of which has a \NO\ solitonic flux tube.\cite{DS}
These strings carry quantum numbers in the group
$\pi_1[U(1)^{N-1}]=[\ZZ]^{N-1}$, {\it not} in $\ZN$!  That is, each of
the $N-1$ Nielsen-Olesen strings carries its own conserved integer
charge. These strings cannot lead to a good model for the dynamics of
$SU(N)$ SYM or YM, which on general grounds must have $\ZN$-carrying
strings.

\begin{figure}
\centering
\epsfxsize=3.5in
\hspace*{0in}
\vspace*{.2in}
\epsffile{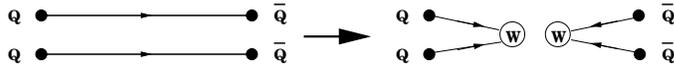}
\caption{A pair of parallel $SU(2)$ strings can break via $W$ boson 
production.}
\label{fig:stringbreak}
\end{figure}

Is this problem serious?  At first glance, the little cheat that I
made just a moment ago rescues the abelian projection.  The magnetic
$U(1)^{N-1}$ theory has a cutoff at the scale $\Lambda$, where its
coupling becomes large.  At that scale, there are massive
electrically-charged gauge bosons of $SU(N)$.  Pair production of
these particles cause certain configurations of parallel strings,
which would naively be stable according to the reasoning of the
previous paragraph, to break.  The charges of these gauge particles
are precisely such that they reduce the conserved symmetry from
$[\ZZ]^{N-1}$ to $\ZN$.  For example, consider the case of $SU(2)$, as
shown in figure \ref{fig:stringbreak}.  An isospin-$\half$ quark and a
corresponding antiquark will be joined by a Nielsen-Olesen string.  This
string is stable.  However, two such quark-antiquark pairs, with
parallel strings, are unstable to reconfiguring their strings via $W$
boson production.  This reflects the claim above that $W$ production
reduces the symmetry under which the $SU(2)$ strings transform from
$\ZZ$ to $\ZZ_2$.

Topologically speaking, all seems well --- the charges seem to be the
expected ones --- but the dynamics of the theory still poses a serious
problem.  Although $[\ZZ]^{N-1}$ is not exactly conserved, it is {\it
approximately} conserved.  To see this in the $SU(2)$ example, note
that the $W$ pair-production requires an energy of order $\Lambda$.
The string tension (its energy per unit length) is $T =
R_{conf}^{-2}\approx \mu\Lambda$ in this theory; here $R_{conf}$ is
the confinement length.  In order for the string to have enough energy
to break, it should have length $L$ such that $TL\sim\Lambda$.  This
implies
\bel{LvsR}
L \geq {1\over\mu} \gg {1\over \sqrt{\mu\Lambda}} = R_{conf} \ ,
\ee
so only enormously long strings can break.  Furthermore, since the
strings' energy density is very low, it takes a large fluctuation to
generate $W$ boson pairs.  This in turn means that the rate for the
transition in figure \ref{fig:stringbreak} is very slow.  In short,
the string pair shown in the figure is {\it metastable} for
$\mu\ll\Lambda$.  The $\ZZ$ symmetry is still approximately conserved
by the dynamics.  Similar arguments apply for $SU(N)$.

\ures The strings of weakly broken \ntwo\ $SU(N)$ SYM carry an approximately
 conserved $[\ZZ]^{N-1}$ symmetry, which contains the expected $\ZN$
 as an exactly conserved subgroup.  This leads to metastable string
configurations not expected in \none\ SYM and in YM.

The physical consequences of this approximately conserved symmetry are
potentially dramatic.  In SYM, a pair of parallel strings which carry
fluxes of charge $k$ and $p$ under $\ZN$ should undergo a rapid
transition to a string carrying charge $k+p$ mod $N$.  (In QCD, this
implies that the string between a quark and an antiquark is the same
as the string between a quark and a diquark.)  But this transition is
inhibited in the broken \ntwo\ gauge theories for small $\mu$. This
implies that numerous, distinct, metastable configurations of strings
may connect a quark in the ${\bf N}$ representation to a corresponding
antiquark.  For example, for $k=1,\cdots N/2$, a pair of parallel
strings, with charges $k$ and $N-k+1$ respectively, carry total charge
$1$; they therefore may, as a pair, join a quark and antiquark.
(Since a string with charge $N$ is no string at all, the $k=1$ case is
the expected one.)  These $N/2$ metastable configurations have
different energies per unit length, and in principle can give
physically distinct quark-antiquark meson Regge trajectories.

Indeed, in \ntwo\ SYM, the dynamics of the theory breaks the Weyl
group, so the $N$ colors of quark are inequivalent.  As shown by
Douglas and Shenker, each color of quark prefers a {\it different}
choice of string pairs.\cite{DS}  This leads to their most surprising
conclusion.

\ures In weakly broken \ntwo\ \susic\ $SU(N)$ gauge theory, the
quark-antiquark mesons exhibit $N/2$ Regge trajectories,\cite{DS}
instead of one as expected in \none\ SYM and in YM.

What happens to these extra trajectories as $\mu\rarr\infty$ and the
theory approaches pure \none\ SYM?  As can be seen from \Eref{LvsR},
the obstructions to $W$ pair production go away as $\mu\rarr\Lambda$.
(Note the formulas which lead to \eref{LvsR} receive corrections at
order $\mu/\Lambda$.)  The extra Regge trajectories become highly
unstable and disappear from the spectrum.\cite{DS} There is no sign of
conflict with the usual SYM expectations of a single Regge trajectory
and of strings with a $\ZN$ symmetry.  However, the $U(1)^{N-1}$
magnetic theory which we used to describe the weakly broken \ntwo\
theory becomes strongly coupled in this limit, and so one cannot study
this picture quantitatively.

In summary, although broken \ntwo\ \susic\ gauge theory can be used to
show that \none\ SYM is a confining theory,\cite{nsewone,sun,DS} it is
not a good model for the glueballs and heavy-quark hadrons of \none\
SYM.  This is a direct consequence of the dynamical abelian
projection, which leads to an abelian dual description.  The
condensation of its abelian monopoles leads to confinement by
Nielsen-Olesen strings, which carry (approximately) conserved integer
charges that (S)YM strings do not possess.  These charges alter the
dynamics of bound states, leading to a spectrum and to hadron-hadron
interactions very different from those expected in SQCD and found in
QCD.  By contrast, these problems are avoided in the broken \nfour\
description of confinement given in sections 5.5-5.6.

{\underline {Moral:}} The use of abelian projection, and the
construction of a dual abelian gauge theory, has inherent difficulties
in explaining the dynamics of YM strings.  One must therefore
use abelian projection with caution.  It may provide good answers for
a limited set of questions, but for other questions it may fail badly.

\subsection{String tensions in $SU(N)$}

The discussion to this point has been entirely
qualitative.  Are any quantitative predictions possible?

A very useful theoretical quantity to study is the ratio of tensions
of strings carrying different charge under $\ZN$.  A confining string
of $SU(N)$ SYM or YM with quantum number $k$ under $\ZN$ has a
tension $T_k$ which depends on $k,N$ and the strong coupling scale
$\Lambda$.  On dimensional grounds $T_k = \Lambda^2 f(k,N)$.  While no
analytic technique is likely to allow computation of $T_1$, it is
possible that $T_k/T_1$ can be predicted with some degree of accuracy.
Note that charge conjugation implies $T_k = T_{N-k}$, so $SU(2)$ and
$SU(3)$ have only one string tension.  We must consider $SU(4)$ and
higher for this to be non-trivial.

While the ratio of tensions cannot be computed in continuum SYM or
YM, it has been calculated in theories which are believed to be in
the same universality class.  These theories often have multiple mass
scales $\mu_i$ and thus in principle it may be that $T_k = \Lambda^2
h(\mu_i/\Lambda,k,N)$.  However, in all cases studied so far, it has
been found that $T_k = g(\mu_i,\Lambda) f(k,N)$, where $f$ is
dimensionless, and $g$ is a dimensionful function which is independent
of $k$.  Note $g$ cancels out in ratios of tensions.  Thus our attention may
focus on the dimensionless function $f(k,N)$ as a quantity which may
be compared from theory to theory.

Some previous calculations include the well-known
strong-coupling expansion of YM, which to leading order
gives
\bel{fsc}
f_{sc}(k,N) \propto k(N-k) \ ,
\ee
and the results of Douglas and Shenker for weakly
broken \ntwo\ gauge theory~\cite{DS}
\bel{fds}
f_{DS}(k,N) \propto \sin{\pi k\over N} \ .
\ee
The considerations of sections 5.5-5.6 suggest another calculation: if
the string solitons in broken \nfour\ $SU(N)$ gauge theory were
computed, the ratios of their tensions would be of considerable
interest.  At the time of writing these soliton solutions have not
appeared in the literature.

There is one other technique by which string soliton tensions may be
computed, using M theory!  M theory is eleven-dimensional supergravity
coupled to two-dimensional membranes.  The theory also has
five-dimensional solitons, called ``five-branes'', whose worldvolume
is six-dimensional.  The theory on the worldvolume of the five-brane
is poorly understood but is known to be a 5+1 dimensional conformal
field theory.  One can construct five-branes with rather strange
shapes whose worldvolume theory contains, at low-energy, a sector with
the same massless fields and interactions as \ntwo\ $SU(N)$ SYM, or
broken \ntwo\ $SU(N)$ SYM, or pure \none\ SYM, or even (in principle) pure
non-supersymmetric YM.  (These theories also contain an infinite tower
of massive particles, all neutral under $\ZN$; thus they are
potentially in the same universality class as, but should not be
confused with, the gauge theories we are interested in.\cite{witMb})
Witten showed membranes can bind to these five-branes, making objects
that carry a $\ZN$ charge and look in 3+1 dimensions like
strings.\cite{witMb} It was then shown~\cite{hsz} that these strings
indeed confine quarks and reproduce the results of Douglas and Shenker
in the appropriate limit.  However, the string tension ratios can be
computed (naively) even when the \ntwo\ breaking parameter $\mu$ is
large. One finds that the tensions are given by~\cite{hsz}
\bel{TM}
T_k = g(\mu,\Lambda,R_0) f_{DS}(k,N)
\ee
where $g$ is a complicated dimensionful function which cancels out of
tension ratios, and $f_{DS}$ is as in \Eref{fds}.
Thus, M theory suggests that the tensions ratios satisfy \eref{fds}
for large as well as small breaking of \ntwo\ \susy.
One must be careful with this result, however.\cite{hsz,esbrane}
No non-renormalization theorem
protects the result \eref{TM} when the \ntwo-breaking parameters are large.
The overall coefficient function $g$ is certainly 
renormalized.  The question is whether $f$ is
strongly renormalized or not. Only a lattice computation
will resolve this issue.

In summary, we so far have results on confining strings from the
strong coupling expansion on the lattice (which only exists for
non-\susic\ theories), from broken \ntwo\ \susic\ gauge theory in the
continuum, and from the M theory versions of YM and SYM.  There are
a couple of interesting observations worth making.

First, the functions $f_{sc}$ and $f_{DS}$ have different large $N$
behavior.  They agree at leading order, but while the first correction
is at order $1/N$ for $f_{sc}$ (as we would expect for an $SU(N)$
theory), the first correction for $f_{DS}$ is order $1/N^2$!  The fact
that the $1/N$ correction in $f_{DS}$ {\it vanishes} is surprising,
and the physics that lies behind this feature has not been explained.

Second, and more important, each calculable limit gives $T_k$ as a
concave-down function of $k$, with $T_k< kT_1$ for all $k$; thus in
each case a string with charge $k\leq N/2$ is stable against decay to
one or more strings of smaller charge.  In short, flux tubes of small
charge attract one another and form tubes of larger charge.  In all of
these limits, then, the gauge theory is a type I dual superconductor.
I believe this result is robust and will be confirmed numerically in
both SYM and YM.  Specifically, I personally expect that
$2T_1>T_2>T_1$ for $SU(N)$, $N\geq 4$.  The reasoning for this is the
following.  For $N=2$, we have $T_2=0$, while for $N=3$ we have
$T_2=T_1,T_3=0$ on general grounds.  For large $N$, we would expect
$T_k= kT_1 \pm \order(1/N)$ since a $k$-string is a bound (or unbound)
state of $k$ $1$-strings, but string-string interactions are of order
$1/N$.  Any reasonable interpolating formula will satisfy
$kT_1>T_k>T_1$ for $N>4$, $1<k<N-1$.  Inspired by this argument, Ohta
and Wingate~\cite{OW} have compared $T_1$ and $T_2$ in $SU(4)$ by
computing the potential energy between sources in the ${\bf 4}$ and
${\bf\bar4}$ representation ($V_{4\bar4}(r) \sim T_1r$) and the energy
between sources in the ${\bf 6}$ representation ($V_{66}(r) \sim T_2
r$.)  Their results suggest indeed that $2T_1>T_2>T_1$.

\subsection{Discussion}

What has been obtained here?  Two qualitatively different descriptions
of confinement have been found, and neither can be continued directly
to the theory of interest.  One looks similar to abelian projection,
while the other absolutely does not.  What are we to make of this
situation, and how are we to reconcile apparent contradictions?

\begin{figure}
\centering
\epsfxsize=2.5in
\hspace*{0in}\vspace*{0in}
\epsffile{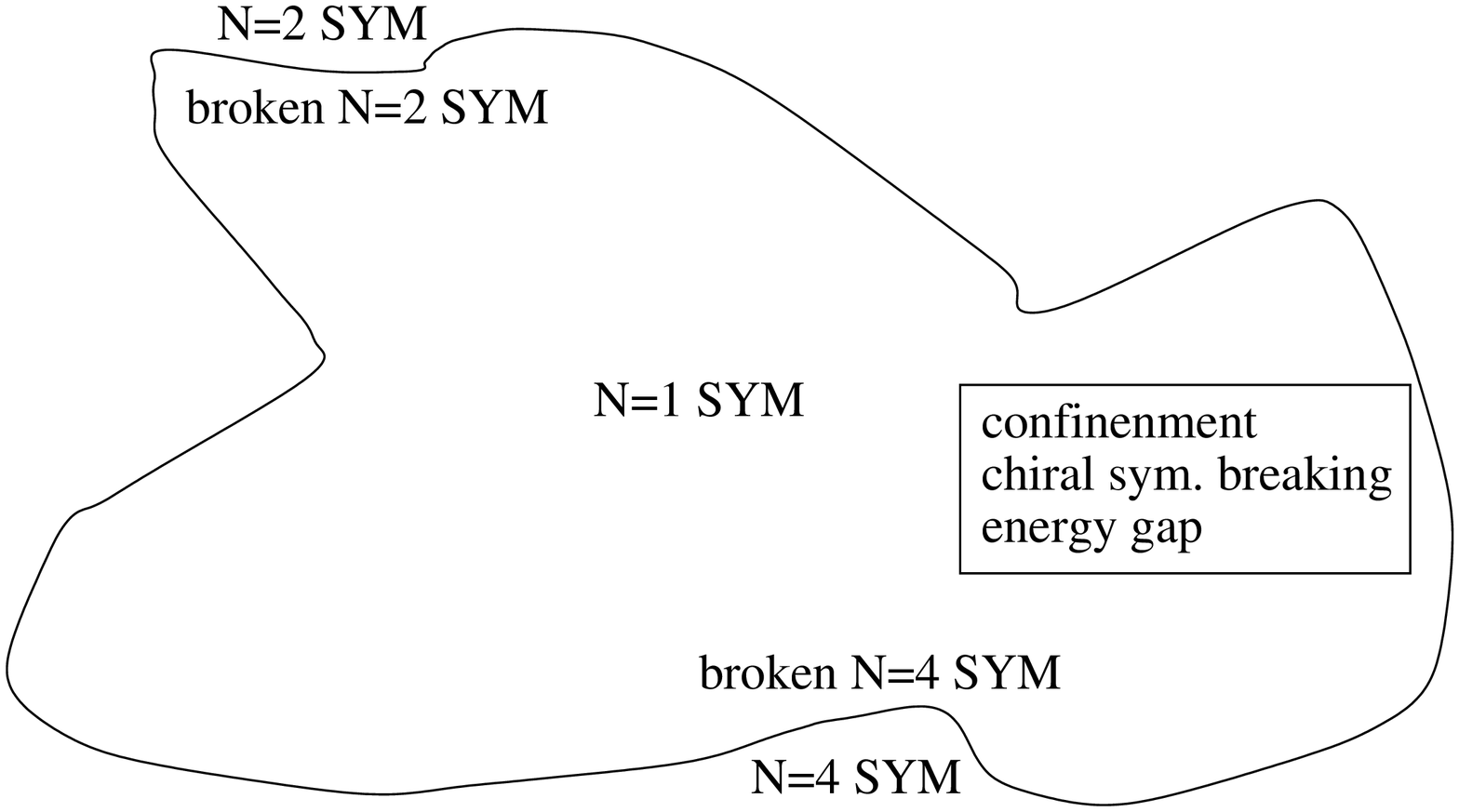}
\caption{The universality class containing \none\ SYM,
with \ntwo,\nfour\ SYM lying just outside.}
\label{fig:thyspace}
\end{figure}

I believe the correct way to view this situation~\footnote{I thank
N. Seiberg for discussions on this point.} is the following.  Consider
the space of theories in the same universality class as SYM,
\FFig{thyspace}.  Although all of these have a gap, confinement, and
chiral symmetry breaking, only theories near a phase boundary, at the
edge of the space, may be expected to have a weakly-coupled
Landau-Ginsburg-type description.  These dual descriptions may be used
to establish the universal properties of \none\ SYM.  Theories far
from the boundary, such as \none\ SYM itself, may simply not have any
such description, and so there may not be any weakly-coupled effective
theory for describing the string charges, hadron spectrum, and
confinement mechanism of SYM. The same logic may apply to non-\susic\
YM.\footnote{Note that it is very helpful, in the supersymmetric case,
that the vanishing energy of the vacuum and holomorphy in the
parameters of the Lagrangian ensure that there are no phase boundaries
other than the one I have drawn in \FFig{thyspace}.\cite{powerholo}
Whether this is true in the non-supersymmetric case must be tested
numerically.}

In the end, then, our conclusions are very weak --- we can show
SYM is confining, but we cannot really say much about the mechanism
which confines it.  Any monopole description of
confinement in SYM or YM is likely to be strongly coupled.  Is such a
description useful? or unambiguous?  It seems unlikely to be
predictive, in any case. It may be disappointing, but it appears
likely there is no simple magnetic description of confinement in
nature.

\section{Connecting to Non-Supersymmetric Theories}

Most physicists outside of the field of \susy\ are inclined to think
of supersymmetric theories, especially those with extended
supersymmetry, as esoteric curiosities with no real importance for
physics.  I now intend to convince you that this is far from the
truth.  In fact, I will now argue that there is a direct link between
the spectacular properties of \nfour\ SYM and the properties of
ordinary non-\susic\ YM.  I will then show a somewhat more qualitative
connection between SQCD and ordinary QCD. 
\subsection{Linkage of YM to \nfour\ SYM}

\begin{figure}
\centering
\epsfxsize=2.7in
\hspace*{0in}\vspace*{0in}
\epsffile{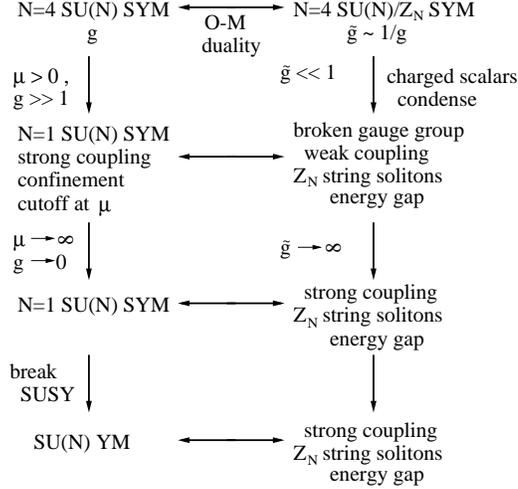}
\caption{The connection of Olive-Montonen duality
to confinement in Yang-Mills theory.}
\label{fig:linkageA}
\end{figure}

Consider the linkage diagram in \FFig{linkageA}.  We begin at the top
with \nfour\ $SU(N)$ SYM theory, a conformal field theory with a gauge
coupling $g$. Under Olive-Montonen duality, this theory is rewritten,
using magnetic variables, as \nfour\ $SU(N)/\ZN$ SYM, with gauge
coupling $\tilde g\propto 1/g$.  Next, as discussed earlier, we take
$g\gg 1$ and break \nfour\ \susy\ to \none\ by adding finite masses
$\mu$ for some of the fields.  The resulting theory is strongly
coupled, confines, and has chiral symmetry breaking, as is easily seen
using the weakly-coupled magnetic variables, where the effect of the
\susy\ breaking is to cause the light monopoles to condense, breaking
the gauge group completely, and leading to $\ZN$-carrying solitonic
flux tubes.

The following step is to take $g$ small and $\mu$ to infinity, holding
the strong coupling scale fixed.  The theory becomes pure \none\ SYM
in this limit, and maintains confinement and chiral symmetry breaking
since it remains in the same universality class.~\footnote{See the
previous footnote.}  The magnetic variables become strongly coupled as
$g\rarr 0$, but we expect the $\ZN$ solitonic strings will survive,
since their existence is a consequence of the topology of the
$SU(N)/\ZN$ gauge group.

The last step is to break \none\ SYM to YM.  Here I must assume that
there are no sharp transitions between these two theories --- an issue
which can and should be addressed on the lattice --- in
order to make the linkage complete.  However, because SYM and YM share
many properties, such a conjecture is quite plausible.  If the
transition between these two theories is smooth, then confinement and
the $\ZN$ flux tubes will survive from one theory to the other.

In summary, modulo the conjecture that \none\ SYM and YM are
continuously connected, the specific structure of duality in \nfour\
SYM theories is directly related to --- perhaps even implies --- the
fact that YM is a confining theory with $\ZN$ flux tubes.

\subsection{Linkage of QCD to Duality in \ntwo, \none\ SQCD}

\begin{figure}
\centering
\epsfxsize=3.2in
\hspace*{0in}\vspace*{0in}
\epsffile{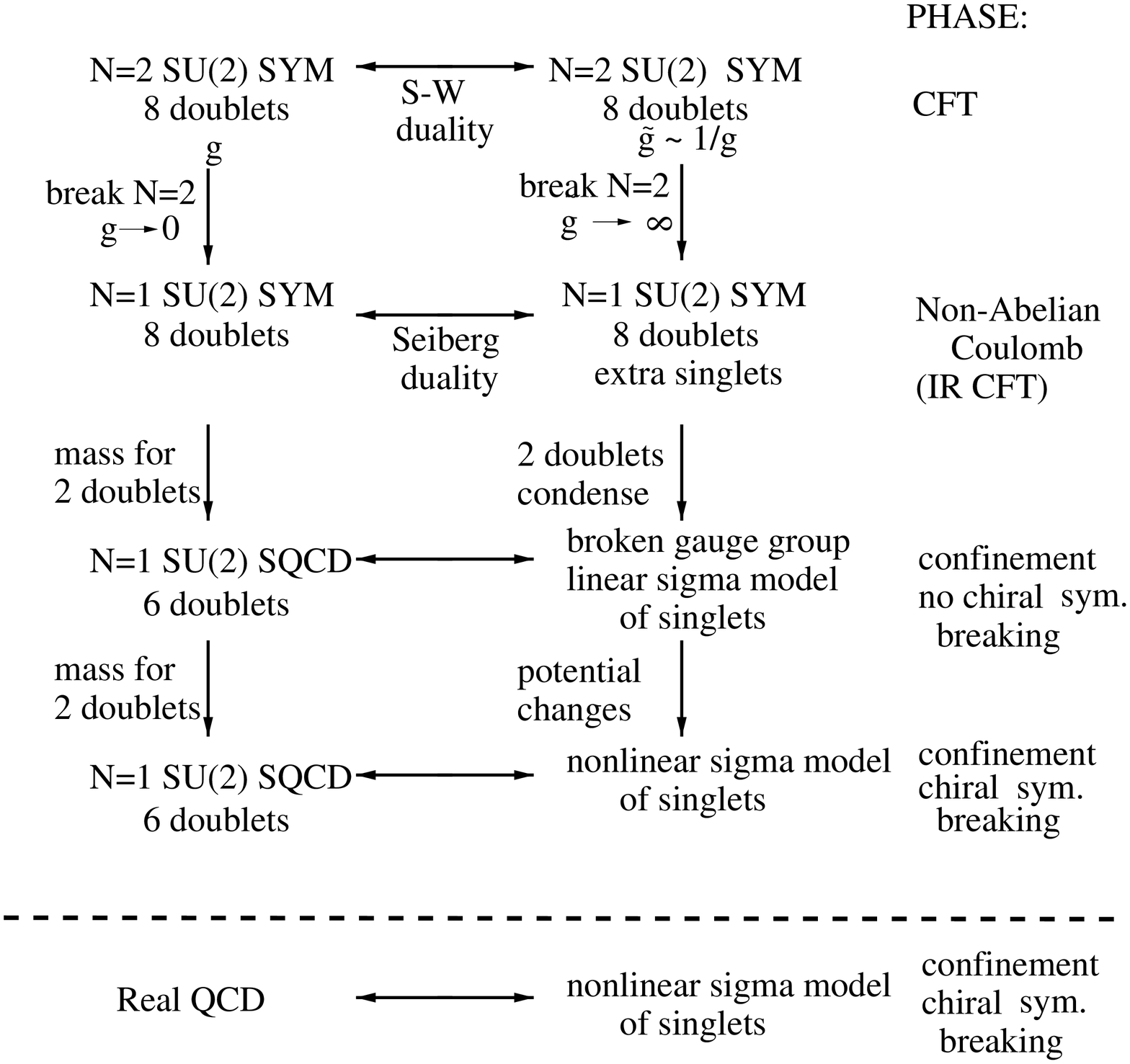}
\caption{The connection of Seiberg-Witten and Seiberg duality to
confinement, chiral symmetry breaking, and non-linear sigma models in
QCD.}
\label{fig:linkageB}
\end{figure}

Now I turn to my second linkage diagram, \FFig{linkageB}, which
relates the duality of finite \ntwo\ theories to that of \none\
theories and then to the properties of real QCD.

At the top of the diagram, we have a finite --- and therefore
conformal --- \ntwo\ SQCD theory, with gauge group $SU(2)$, gauge
coupling $g$, and eight quarks and squarks in the doublet
representation.~\footnote{The choice of $SU(2)$ is for simplicity only;
the same physics applies with slight modification for any $SU(N)$.}
As in the case of \nfour\ SYM discussed earlier, this theory has a
representation in terms of magnetic variables as another \ntwo\ SQCD
theory, which in this case has the same gauge and matter content as
the electric variables but has coupling constant $1/g$.  This duality
transformation was discovered by Seiberg and Witten in their famous
paper of 1994.\cite{nsewtwo}

Now let us break \ntwo\ \susy\ to \none\ by giving mass to the extra
fields in the \ntwo\ gauge multiplet.  The theory becomes \none\ SQCD
with gauge group $SU(2)$ and eight quarks and squarks in the doublet
representation.  This theory has a running coupling, but flows to a
conformal fixed point in the infrared --- it is in the non-abelian
coulomb phase.  The breaking of \ntwo\ \susy\ causes the magnetic
theory to flow to an \none\ SQCD theory with the same charged matter
content but with extra gauge singlets and interactions,\cite{emop}
precisely those required by Seiberg's \none\ duality
transformation.\cite{NAD} In other words, the Seiberg-Witten duality
transformation of the \ntwo\ theory flows to the Seiberg duality
transformation of \none\ SQCD.

Now add masses for two of the electric doublets, leaving a theory of
$SU(2)$ with six doublets.  This causes some magnetic squarks to
condense, breaking the $SU(2)$ magnetic gauge symmetry and leaving a
theory of massless gauge singlet fields $M$. These singlets are
precisely the mesons of the electric variables.  Since the magnetic
gauge symmetry is broken, electric charge is confined.  Thus,
confinement proceeds through a non-abelian generalization of the dual
Meissner effect, and the low-energy magnetic theory --- an
infrared-free non-renormalizable theory with a cutoff --- is the sigma
model describing the confined hadrons.  Examination of the sigma
model, in particular the potential energy $V(M)$, shows that the the
theory has a vacuum in which chiral symmetry is unbroken.

Adding masses for two more doublets merely causes the potential $V(M)$
to change in such a way that there is no longer a
chiral-symmetry-preserving vacuum.  Thus, $SU(2)$ SQCD with four
doublets confines and breaks chiral symmetry.  Shifting to the true
vacuum and renaming the fields as representing fluctuations around
that vacuum, we may rewrite the theory as a non-linear sigma model of
pions and their superpartners.

To go from here to real non-\susic\ QCD is a bit more of a stretch
than in the previous linkage diagram, because the removal of the
scalar squarks from the theory is rather more delicate and much less
well understood.  Rather than raise those questions, I leave the last
step in the diagram as more of a heuristic one.  It is evident that
the duality in \none\ $SU(2)$ SQCD with four doublets, relating the
gauge theory of gluons, quarks and their superpartners to a non-linear
sigma model of pions, is remarkably similar to the transformation
between real-world QCD and the chiral Lagrangian that we use to
describe its infrared physics.  Indeed it is completely justified to
call this QCD-to-hadron transformation ``duality''.  As we have seen,
the duality in confining SQCD can be derived from the Seiberg-Witten
duality of a conformal \ntwo\ SQCD theory. Is QCD-pion duality
likewise embedded in a chain of non-\susic\ duality transformations
similar to those in the diagram?  

We are unable at this time to answer this question, even for
non-chiral theories like QCD, because almost nothing is known about
non-supersymmetric gauge theories other than $SU(2)$ and $SU(3)$ YM
and QCD with a small number of flavors.  This is where lattice
QCD comes in, as it is at the present time almost the only tool
available for studying this issue.

\section{Orbifolds?  Here?}

Orbifolds of $SU(N)$ SYM are potentially rich sources of new insights.
To my knowledge the importance of these theories was emphasized only
recently.\cite{jpms,mjsorbi} At the end of this section I will
illustrate this through yet another linkage diagram.

Orbifolds of supersymmetric gauge theories are common in the recent
string theory literature.  A certain ``parent'' gauge theory is
chosen; the orbifold theory is simply given by throwing away all
fields which are not invariant under a discrete subgroup of the gauge
symmetry.  The resulting theory has the remarkable property that at
large $N$ its perturbation series is the same as the parent theory, up
to some simple rescaling of
$N$.\cite{KachSilver,LawNekVaf,BerKakVaf,Schmorbi} This is because the
planar graphs have a very simple property under this
projection.\cite{berjoh} However, it is not proven that the relation
between orbifold and parent holds non-perturbatively in the gauge
coupling.  Note that one cannot argue that non-perturbative effects
are simply suppressed in the large-$N$ limit, as was long ago
suggested from the fact that instanton effects are suppressed.  Such
arguments are clearly wrong in the case of \none\ SYM, where the
gluino condensate does not vanish at large $N$.  In short, the
``orbifold conjecture'' --- the suggestion that orbifold and parent
are related even non-perturbatively --- is a non-trivial one.

For supersymmetric theories, the parent-orbifold connection often
leads to very interesting results.  Non-supersymmetric orbifolds,
however, are usually unnatural.  Most supersymmetric theories have
massless scalar fields, and their orbifolds do as well.  For
non-supersymmetric orbifolds, the scalar fields remain massless only
to leading order in $1/N$.  At higher order their masses have the
usual quadratic divergences, which although multiplied by $1/N$ are
still divergent!  In other words, the limit in which the cutoff is
taken to infinity does not commute with the infinite $N$ limit.  Thus,
all models of this type with large but finite $N$ are fine-tuned.
(Interesting observations have been made concerning these theories
nonetheless,\cite{KachSilver,Schmorbi} although to what end I am
unsure.)  The important exception, however, is for orbifolds of SYM,
which like SYM have no scalar fields, and which have fermions that are
strictly massless due to chiral symmetries.  These theories are
natural and exhibit many phenomena that we would like to study.

I will focus exclusively on the non-supersymmetric $\ZZ_p$ orbifolds of
SYM.  These have $p$ $SU(N)$ gauge factors, which I will label
$SU(N)_1$, $SU(N)_2$, etc.  There are also $p$ Weyl fermions
$\psi_i$, $i=1,\dots,p$, cyclically charged: $\psi_i$ is charged
under two gauge groups, as ${\bf N}$ under $SU(N)_i$ and as
${\bf\overline N}$ under $SU(N)_{i+1}$.  Like SYM, these theories also
have discrete $\ZZ_{2N}$ global symmetries which can be broken by
fermion condensates.  They also have a $\ZZ_N$ center under
which flux tubes can be charged.~\footnote{For the experts: I am
ignoring the $U(1)$ factors that typically would arise in an orbifold.
These are infrared free and cannot affect the non-perturbative physics
in the infrared.  If necessary they can easily be accounted for using
perturbation theory, as with QED when discussing its effects on QCD;
but they represent a $1/N$ effect in any case.}

The orbifold conjecture applies only when all gauge coupling constants
are precisely equal.  However, let us consider these theories with
arbitrary gauge couplings, understanding that the relation with the
parent theory only holds quantitatively when the couplings are exactly
equal.  This allows us to do a qualitative analysis of the infrared
behavior of the theory.\cite{georgi} In particular, if there is a
hierarchy in the gauge couplings, one can see, from the known physics
of QCD, that these theories exhibit a form of tumbling. Suppose
$SU(N)_2$ has the largest of the couplings.  It will be the first,
then, to become strongly coupled, with all the other gauge groups
acting as weakly-coupled spectators to its dynamics.  From QCD with
three light flavors, we expect $SU(N)_2$ will confine and induce a
condensate for $\psi_1\psi_2$.  Just as in QCD, this condensate breaks
$SU(N)_1\times SU(N)_3$ down to a diagonal subgroup $SU(N)_D$; in the
breaking, the pions of $\psi_1\psi_2$ are eaten by the broken vector
bosons.  The combination of confinement of $SU(N)_2$ and breaking of
$SU(N)_1\times SU(N)_3\rightarrow SU(N)_D$ leaves $p-2$ gauge factors
with $p-2$ Weyl fermions cyclically charged as before.  In short, the
$\ZZ_p$ orbifold tumbles dynamically to the $\ZZ_{p-2}$ orbifold.  The
next stage of confinement and chiral symmetry breaking removes two
more factors, and so on.

If $p$ is odd, then the endpoint of this tumbling is simply the $p=1$
case, which is just \none\ SYM.  In this case the low-energy effective
theory has supersymmetry, as an {\it accidental} symmetry!  The final
step in the dynamics will involve SYM physics: confinement with a
breaking of the discrete chiral symmetry, leaving a set of $N$ vacua
which can have domain walls between them.  If there are no phase
transitions between this regime and the regime with all gauge
couplings equal, then the orbifold limit will have the same
properties.  Thus its seems that the non-perturbative dynamics of the
$SU(N)^p$ theory may indeed closely resemble those of an $SU(N)$ SYM
theory.  This make it seems likely --- though difficult to verify ---
that when $N$ is large the relation between SYM and its orbifolds with
$p$ odd goes well beyond perturbation theory in $g$.

Note also that the $p=3$ case is especially amusing.  Take $N=3$; then
our theory is QCD with three massless quarks, with the flavor symmetry
$SU(3)_L\times SU(3)_R$ gauged and with an additional fermion $\chi$
in the $(3,\bar 3)$ of $SU(3)_L\times SU(3)_R$ to cancel gauge
anomalies.  The confinement and chiral symmetry breaking of ordinary
QCD would make $\chi$ into an $8$ (plus a decoupled singlet) of the
remaining $SU(3)$ diagonal flavor symmetry --- and thus into an
effective gluino of the low-energy theory.

What if $p$ is even?  Then the endpoint of the tumbling is the $\ZZ_2$
orbifold of SYM, which is especially interesting.  It contains
$SU(N)\times SU(N)$ with a {\it Dirac} fermion $\psi$ in the (${\bf
N},{\bf \overline N}$) representation.  Note, however, that the mass
of the fermion must be zero, because of a $\ZZ_N$ discrete axial
symmetry.  For $N=3$, this is just QCD with three light quarks, with
the diagonal subgroup of the chiral $SU(3)_L\times SU(3)_R$ flavor
symmetry gauged.  Thus, we expect this theory to have confinement and
chiral symmetry breaking also, at least when one of the couplings is
taken very small.  Note that unlike QCD, however, the second $SU(N)$
ensures that there really is confinement here, with flux tubes that do
not break and carry a $\ZZ_N$ charge.  For example, sources in the
(${\bf N},{\bf 1}$) representation cannot be screened by the light
matter.  Also, instead of a continuous global symmetry and pions, as
in QCD, the second $SU(N)$ breaks the global symmetry explicitly,
making the pions pseudo-Goldstones.  The only remaining symmetry is
discrete, and we expect it to be broken by a $\bar \psi\psi$
condensate, giving $N$ degenerate vacua.  Thus, like SYM, it will have
domain walls.

The implications of this are exciting.  What makes this orbifold
theory special is that it can be simulated on the
lattice with relative ease.

1) It can be studied when one of its gauge couplings is much
larger than the other. In this case it should look like QCD with
$SU(N)$-diagonal gauged, with the pions as a light charged adjoint, 
and with a secondary, low-energy confinement scale with associated
flux tubes.

2) It can be studied when the gauge couplings are nearly equal.  In
this case, according to the orbifold proposal, many of its properties
should look very similar to SYM.  In the limit where the couplings are
exactly equal and $N$ is taken moderately large, it should become
quantitatively similar to SYM.  For instance, the hadron spectrum should
have surprising degeneracies. Furthermore, the relative
tensions of domain walls should agree with those of SYM, where the
domain walls are believed to be BPS-saturated and the ratios of their
tensions are therefore {\it exactly} predicted!  {\it These relations
allow a quantitative test of the orbifold conjecture beyond
perturbation theory.}  Furthermore, if the orbifold conjecture is
confirmed, {\it this theory gives quantitative tests of predictions in
\none\ SYM, even without simulating SYM on the lattice.}  It could
therefore be used to resolve some of the puzzles still surrounding
non-perturbative aspects of SYM.

3) Finally, the orbifold conjecture should survive quantitatively even
when one compares broken \none\ supersymmetry (with a mass for the
gluino) to this $SU(N)\times SU(N)$ theory with a mass for the Dirac
fermion.  Thus, the $\ZZ_2$ orbifold is potentially a rich opportunity
for studying various non-perturbative phenomena of QCD, SYM and YM
simultaneously, through a theory which essentially is expected to
interpolate between them.  At worst, one will disprove the orbifold
conjecture at the non-perturbative level; at best, one will both verify
it and put it to quantitative use!

\begin{figure}
\centering
\epsfxsize=2.8in
\hspace*{0in}\vspace*{0in}
\epsffile{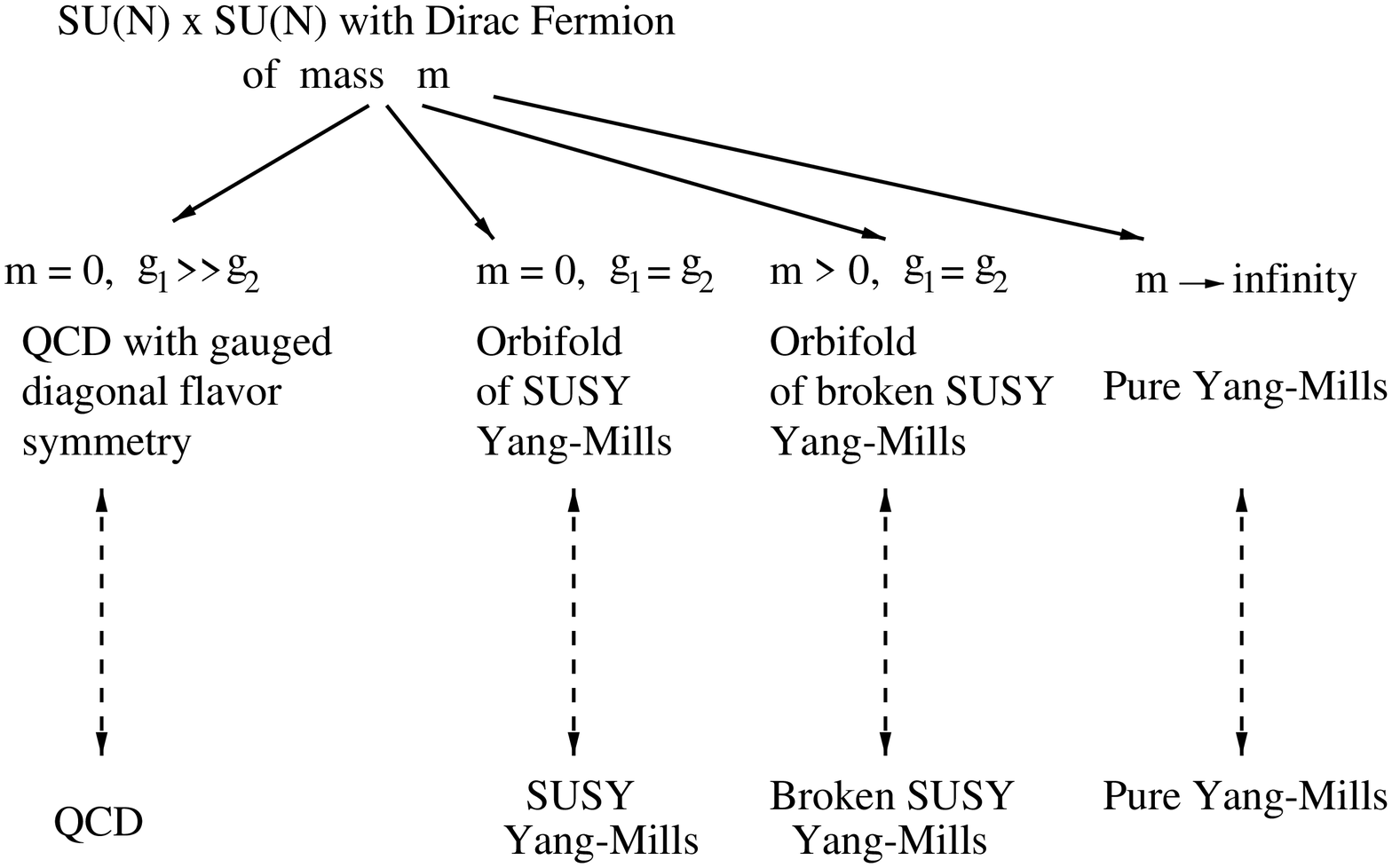}
\caption{The $\ZZ_2$ orbifold of \none\ SYM may be useful for the
study of a number of interesting questions, as it has qualitative and
quantitative connections with several important theories.}
\label{fig:orbilink}
\end{figure}

The interesting properties of the $\ZZ_2$ orbifold of \none\ SYM
are summarized in a third linkage diagram, given in \FFig{orbilink}.

\section{Large $\nc$ Gauge Theory and String Theory}

Now I want to turn to some even more spectacular developments
of the last couple of years.  I will be brief, but I hope
to convey some of the key ideas nonetheless.

The fact that gauge theory, in the limit of a large number of colors,
is in some way connected with string theory, with $1/N$ playing the
role of the string coupling, was first noted twenty-five years ago by
't Hooft.  While many have attempted to make progress in
understanding gauge theories by studying this limit further, a
quantitative approach to $SU(N)$ YM or QCD using $1/N$ as an expansion
parameter has been stymied by the difficulty of determining the
classical string theory which should appear in the $N\rarr\infty$
limit.

Let us consider some obvious facts about this string theory, and the
gauge theories which might be studied using this approach.  First,
critical (that is, $d=10$) superstrings have gravity and
supersymmetry.  Since QCD has neither a spin-two hadron with only two
polarization states (a graviton) nor supersymmetry, clearly its string
theory is unrelated to critical superstrings.  A second obvious fact
is that only confining theories have physical string-like flux tubes
and associated area laws for Wilson loops.  Other theories, including
conformal field theories such as \nfour\ SYM, have no physical
string-like behavior and should not be associated with string
theories.

We will not get anywhere using these obvious facts, however, because
they are wrong.  Following on the work of Gubser and
Klebanov,\cite{gubkleb} and followed in turn by work of those authors
with Polyakov,\cite{GKP} of Witten,\cite{ewAdS} and then of a flood of
others, Maldacena proposed a bold conjecture~\cite{maldacon} relating
supersymmetric conformal field theories in four dimensions to
superstring theory.  I will now state this conjecture, defining my
terms as I go.

\subsection{Maldacena's Conjecture}

According to Maldacena's idea, \nfour\ SYM theory with $N$ colors and
coupling $g$ is related to Type IIB superstring theory (a
ten-dimensional theory of closed strings with IIB supergravity as its
low-energy limits --- its massless fields are a graviton $G_\mn$,
antisymmetric tensor $B_\mn$ and dilaton $\phi$ along with
``Ramond-Ramond'' 0-index, 2-index and 4-index antisymmetric tensor
fields $\chi, A_\mn, C_{\mu\nu\rho\sigma}$.)  The string theory exists
on a ten-dimensional space consisting of a five-sphere
($x^2_1+x^2_2+x^2_3+x^2_4+x^2_5+x^2_6=R^2$, a space of constant
positive curvature) times a five-dimensional Anti de Sitter space
($-x^2_1-x^2_2+x^2_3+x^2_4+x^2_5+x^2_6=R^2$, a space of constant
negative curvature) with non-zero flux of $\partial_\kappa
C_{\mu\nu\rho\sigma}$.  The radius of the sphere and of the AdS space
are both $R\propto(g^2N)^{1/4}$, so the curvature of the space is
small at large $g^2N$. The string coupling $g_s$ is the square of the
gauge coupling $g_s\propto g^2\propto R^4/N$.  (Notice the $N$
dependence accords with 't Hooft's original observation.)

 Where is the four-dimensional gauge theory in this ten-dimensional
string?   The five-dimensional AdS space
has a four-dimensional boundary, and it is there that the gauge theory
is to be found.  Note that it has long been understood from the work
of Polyakov that non-critical strings dynamically grow an extra
dimension, so the presence of a five-(plus-five)-dimensional string
theory in the context of a four-dimensional gauge theory is perhaps
not so shocking.  What is astonishing is that the string theory
involved is the well-understood critical superstring, and that it is
believed to be {\it equivalent} to the gauge theory on the boundary.

\begin{figure}
\centering
\epsfxsize=1.6in
\hspace*{0in}\vspace*{0in}
\epsffile{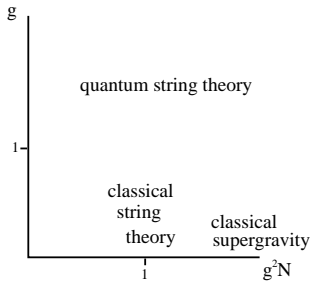}
\caption{The conjectured duality between superstring theory and gauge
theory connects gauge theory to classical superstring theory at small
$g$ and further to classical supergravity at large $g^2N$.}
\label{fig:AdScplg}
\end{figure}
This conjecture has been extremely well-tested in the large $N$, large
$g^2N$ regime, in which the full quantum string theory reduces to
classical (small $g_s$) supergravity (small curvature), as shown in
\FFig{AdScplg}.  The symmetries and operators of the
two theories match;~\cite{maldacon} and the baryon operator (which is
non-perturbative in string theory since $N\sim 1/g_s$) has been
identified with a D-brane (a soliton).\cite{ewDbary} Furthermore, the
Wilson loop of the gauge theory has been identified with the boundary
of a string worldsheet in the larger space.\cite{maldaloop,reyloop}
This makes it possible to explain how this conformal \nfour\ SYM
theory can be stringy: although the Wilson loop is the boundary of a
string, the string does not live on the boundary but hangs into the
bulk, and so the value of the Wilson loop as a function of its size
{\it depends on the geometry outside the boundary.}  The difference
between area law and perimeter law in various gauge theories thus is
translated into differing spatial geometries in their string-theory
duals.

These ideas have been further extended in a number of directions.  In
particular, it is easy to study finite temperature, and to use the
fact that high-temperature five-dimensional SYM has infrared behavior
equivalent to four-dimensional non-\susic\ YM at strong
coupling.\cite{ewThermo,reyThermo,BISYThermo} It is straightforward to
show that this strongly coupled theory confines.  It is also possible
to compute its spectrum of glueballs.\cite{AdSglubl}  Remarkably, the
ratios of certain glueball masses match rather well to lattice results
in ordinary QCD away from strong coupling, both in three and four
dimensions.  However, although this is surprising and possibly an
interesting statement about this particular strong-coupling limit,
there is no clear reason for great excitement.  There are many extra
states in the spectrum which do not arise in YM theory; the glueball
mass is not naturally related to the string tension; and there is no
systematic approach toward recovering real YM theory starting from
this limit.

Other extensions include adding matter (which leads to both open and
closed strings), reducing supersymmetry, changing the number of
dimensions, and studying non-conformal theories at zero and finite
temperature.  Various expected phase transitions, such as
deconfinement at high temperature, have been observed.

Recent work~\cite{jpms,ikms,maldanunez,vafaDsix} has brought us much
closer to a study of \none\ SYM, its QCD-like orbifold, and pure YM.
These papers have shown several different methods for obtaining
gravitational descriptions of theories in the same universality class
as SYM.  For example, one of these~\cite{jpms} gives the stringy
description of the breaking of \nfour\ SYM, as discussed earlier and
illustrated in \FFig{linkageA}.  All the properties of SYM ---
confinement, flux tubes, domain walls, baryon vertices, chiral
symmetry breaking --- have been identified as coming from geometry and
from various branes and strings in various interesting topological or
dynamical configurations.  It is truly remarkable how the physics of
strings and branes in a particular ten-dimensional spacetime manages
to reproduce so much of the physics of SYM, and, when heavy quarks are
added, of a theory resembling QCD.  But this is an article in and of
itself.  I leave this topic for the future.

It should still be noted, however, serious obstacles lie in the path
of any attempt to apply these string theory techniques to true SYM,
ordinary YM or physical QCD, where for any fixed $N$ the value of $g$
runs such that $g^2N$ is not always large and $g$ is not always small.
Where $g^2N$ is small, string theory is not a good description of the
theory; perturbative field theory is, naturally, the only good one.
In the regime where $g^2N$ is of order one, where the hadrons with
mass of order 1 GeV are likely to be found, supergravity is
insufficient; string theory is required for the large $N$ limit.
Unfortunately, almost nothing is known about string theory on an AdS
background with Ramond-Ramond fields, even at the classical level,
since the usual world-sheet formulation of string theory cannot be
easily generalized to this case.  Even were this problem solved, there
is no guarantee that the solution will be easy to use.  And worse, one
must somehow understand quantitatively the transition from the
perturbative gauge theory regime to the stringy regime. I cannot tell
you whether these obstacles will be overcome tomorrow, next year, or
in the fourth millennium; but in any case the difficulties are such
that it seems unlikely these approaches will become a quantitative
competitor to lattice gauge theory in the near future.

Nonetheless, it is remarkable that a sensible and definite proposal
for the large $N$ expansion of gauge theory has been made and has
passed some non-trivial tests --- and that it appears to involve
superstring theory!  And we must certainly ask whether in fact
superstring theory can be {\it defined} using gauge theory.

\subsection{A Final Linkage Diagram}

\begin{figure}
\centering
\epsfxsize=2.8in
\hspace*{0in}\vspace*{0in}
\epsffile{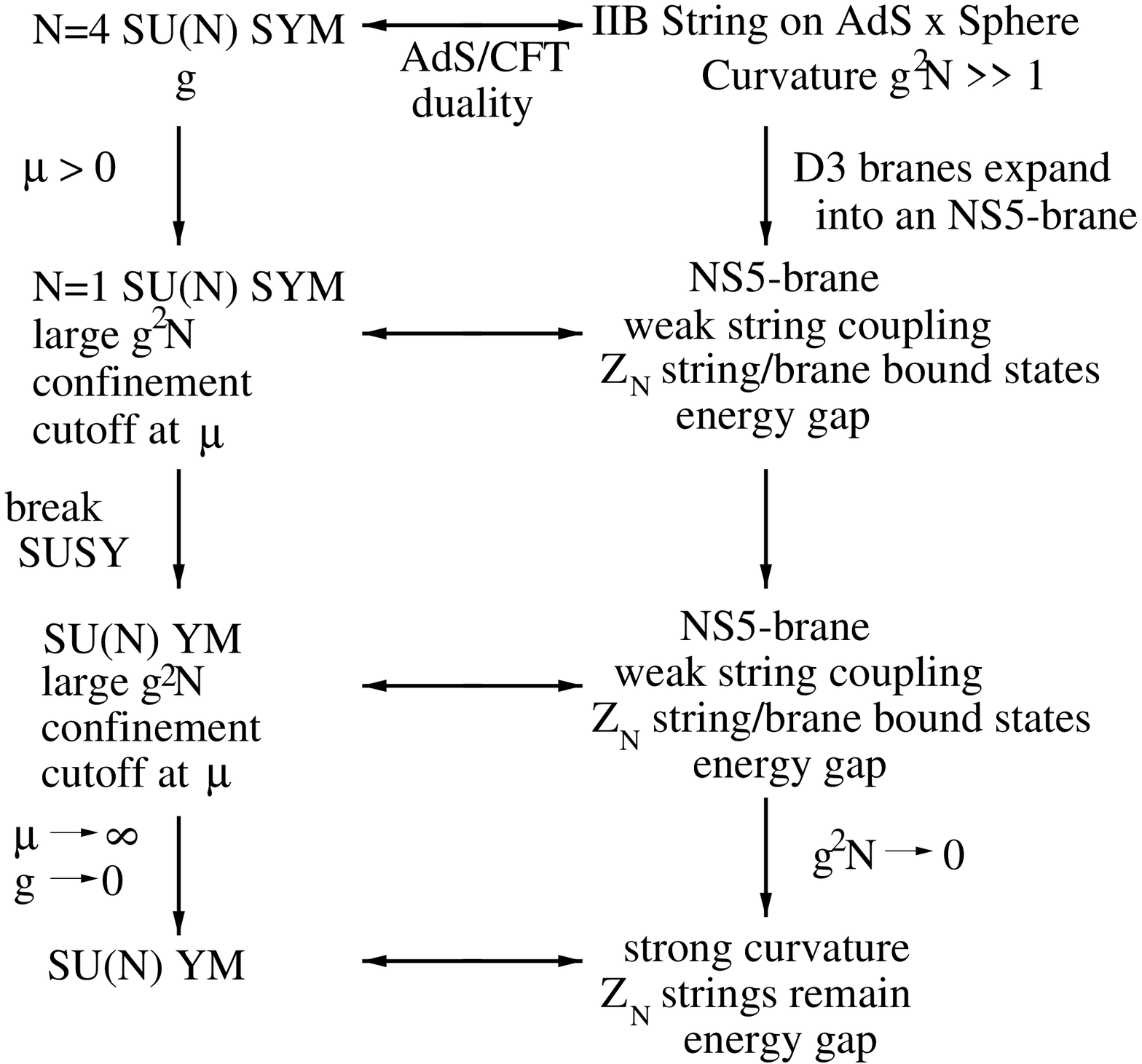}
\caption{The Maldacena conjecture generalized to
the case of \nfour\ SYM broken to \none\ SYM, as
in \FFig{linkageA}.}
\label{fig:linkageC}
\end{figure}

To again bring home how apparently esoteric results on theories with
extended supersymmetry can have implications for real-world physics, I
want to restate using \FFig{linkageC} the connection between QCD and
string theory as presently conjectured and partially understood.  I
must warn the reader that I will be speak rather loosely when
describing this diagram, as this discussion is intended for novices
who want only a rough idea of the physics.  I ask experts to forgive
the obvious misstatements.

Let us begin with \nfour\ SYM and break it to \none, as in linkage
diagram \FFig{linkageA}.  According to the Maldacena conjecture this
theory is dual to a superstring theory which is approximately
$AdS_5\times S^5$ near the boundary of $AdS$ space, but which has an
additional non-zero flux on it corresponding to the \nfour-breaking
masses.  As we move away from the boundary, this new flux grows and begins to react back on the spacetime.
This back-reaction causes a Neveu-Schwarz (NS) 5-dimensional ``brane''
(an object which is dual to strings just as monopoles are dual to
electrons) to appear out of the vacuum.\cite{jpms} It wraps around an
$S^2$ of the $S^5$, as well as filling ordinary three-dimensional
space.  It can be shown that fundamental strings stretched along a
direction of ordinary three-dimensional space can bind to this NS
5-brane.  These strings correspond to electric flux in the gauge
theory.  The fact that they bind to the NS 5-brane, forming a state
of definite tension, corresponds to the
fact that in the SYM theory the flux lines cannot expand to
arbitrarily low density, but instead make flux tubes of
definite tension.  One can also show these string-brane bound states
carry a $\ZZ_N$ charge.  Indeed, the physics of \FFig{linkageA} has
a direct representation in the string theory.

\section{Outlook for QCD}

I will conclude with a short list of important questions raised
in this article, and mention some specific ideas for
future numerical work.

\subsection{Some questions}

First and foremost, what {\it is} duality?  We still have no explicit
understanding of non-trivial duality transformations in three or four
dimensions, and not nearly enough even in two dimensions.  Do we need
a reformulation of field theory itself?  What does duality in string
theory teach us?

Does Olive-Montonen duality imply confinement in non-supersymmetric
Yang-Mills theory by $\ZN$-carrying electric flux tubes?  I
have shown that it does so for \none\ SYM, but one must show there
is no phase boundary separating the two theories. 
What are the $\ZN$ string tensions in pure $SU(N)$ YM?  Do the ratios
of tensions fit any known formulas, such as those of \Eref{fsc} or
\Eref{fds}?  

  What is the phase structure of QCD as a function of gauge group,
matter content, and interactions?  Does QCD have duality similar to
SQCD?  Is there a free magnetic phase? 

Is the orbifold conjecture for large-$N$ field theory correct?  Can it
be verified, using lattice computations, for the QCD-like orbifold of
\none\ SYM described above?  If it works, how can we best put it to
use?  Can the domain walls in the orbifold be quantitatively compared
to those predicted in \none\ SYM?

Can string theory give us a quantitative understanding of \none\ SYM,
and perhaps non-supersymmetric theories as well?  Can it at least
give us the hadron spectrum?  Is the large-$N$ stringy expansion
quantitatively tractable?

We have seen repeatedly that massive matter, when added to a theory,
can make aspects of its physics easier to understand.  What matter (or
additional interactions) might we add to non-supersymmetric QCD to
make some of these questions more accessible either analytically or
numerically?

\subsection{Some proposals for the lattice}

Within this set of questions, there are several projects that
I hope will eventually be undertaken by the lattice community.

1) \underline{The transition from SQCD to QCD}: I hope that it will
soon be realistic to simulate $SU(2)$ or $SU(3)$ QCD with a Majorana
spinor, of mass $m_\lambda$, in the adjoint
representation.\cite{latSQCD} Tuning the mass to zero to make the
theory \none\ SQCD may be difficult, but many properties of the
theory, including its strings, should not be sensitive to $m_\lambda$
as long as it is much smaller than the confinement scale $\Lambda$.
It would be very interesting to map out the behavior of the theory,
including the strings and their tensions $T_k$, as a function of
$m_\lambda/\Lambda$.  As I mentioned earlier, the absence of a
transition in the properties of the strings would establish the
linkage I have proposed between Olive-Montonen duality and the $\ZN$
strings of $SU(N)$ YM.

2) \underline{Ratios of string tensions of YM}: I have argued here
that YM should be a type I superconductor, and that the same should
hold for \none\ SYM and for broken SYM.  This conjecture must be
studied in theories with at least four colors.  It has passed its
first test~\cite{OW} but there is much more to do before it can be
conclusively verified.

3) \underline{$SU(N)\times SU(N)$ with a Dirac fermion
in the $({\bf N},{\bf\overline N})$}: this $\ZZ_2$ orbifold of \none\ SYM
should be quantitatively similar to \none\ SYM in many respects when
$N$ is large.  It would be interesting to study this theory on the
lattice, at least for $N=2$, 3 and 4; with the two gauge couplings
equal, the trend toward a hadron spectrum with surprising degeneracies
should be visible, and the ratio of the domain wall tensions in
$SU(4)$ should be close to that predicted by $SU(4)$ \none\ SYM.  As
emphasized in \FFig{orbilink}, there are several regions of interest
where the theory has relations with pure QCD, pure YM, and broken
\none\ SYM.

4) \underline{Phase structure of QCD}: A more difficult goal is to
fully map out the low-energy phases of non-supersymmetric QCD as a
function of gauge group, matter content, and interactions (including
non-renormalizable ones.)  In \susic\ theories, we have begun to
understand the complicated subject of the long distance physics of
gauge theories as a function of their gauge group $G$, their matter
representations $R_i$, and their interaction Lagrangian (including
non-renormalizable terms.)  Similar information would be welcome in
the non-\susic\ case.  We know that the non-abelian coulomb phase
exists at large $\nc,\nf$ when the one-loop beta function is very
small, but how far does it extend away from this regime?  What
properties does the theory exhibit as it makes the transition from the
perturbative regime to the conformal regime?  The confining phase in
supersymmetric theories is the exception, not the rule; which
non-supersymmetric theories actually confine?  Which ones break chiral
symmetry, and in what patterns?  What are their confining string
tensions $T_k$ and their low-lying hadron spectra?  Effects involving
instantons, fractional instantons, monopoles, {\it etc.} may play an
important role in some theories --- but which ones, and what effects
are they responsible for?  The free magnetic phase may not exist in
non-supersymmetric theories --- perhaps it requires the massless
scalars of SQCD --- but it would be very exciting if it were found
(and a tremendous coup for the group which demonstrates its
existence!)  The existence of this phase would be a sufficient but not
necessary condition for gauge theory--gauge theory duality in the
non-\susic\ context, which could also perhaps be shown in the context
of the non-abelian coulomb phase.  And of course, we must not leave
out the possibility of new exotic effects which do not occur in the
\susic\ case.

It is important to emphasize that these questions are by no means of
purely academic interest.  The problem of electroweak symmetry
breaking has not been solved, and there remains the possibility that
the symmetry breaking occurs through a technicolor-like scenario, in
which it is driven by chiral symmetry breaking in a strongly-coupled
gauge theory.  Technicolor models with physics similar to QCD have
been ruled out by precision measurements at LEP.  However, if the
physics is considerably different from QCD, then we have no predictive
tools, and therefore no experimental constraints.  It will be an
embarrassment to theorists if the LHC discovers evidence of strong
dynamics of a type that we simply do not recognize.  It is therefore
important for model-building and for comparison with experiment that
we improve our understanding of strongly-coupled non-\susic\ theories.

I should mention also that there are potential spinoffs from such a
program in the areas of supersymmetry breaking (which can be detected
but not quantitatively studied using presently available analytic
techniques) and in condensed matter systems where many similar physics
issues arise.

To answer these questions, which require studying theories with very
light fermions and often a hierarchy of physical scales, will require
powerful lattice techniques not yet fully developed. I encourage the
reader to think about how best to pursue this program of
study!~\footnote{A useful testing ground is to be found in
three-dimensional abelian gauge theories, both with and without
supersymmetry.  The \susic\ theories are known to have an intricate
phase structure, with a duality transformation (called ``mirror
symmetry'')~\cite{kinsddd,superQED} and interesting large $\nf$
behavior.  There are examples of non-trivial fixed points,
infrared-free mirror gauge theories, chiral symmetry breaking and
confinement.  Many of these phases are likely to show up in the
non-\susic\ case, and lattice approaches to studying non-\susic\ QCD
could be tested in these theories.}


\section*{Acknowledgements} I am grateful to many colleagues
for conversations, including M. Alford, D. Kabat, R. Leigh, A. Hanany,
K. Intriligator, I. Klebanov, J. March-Russell, J. Polchinksi,
N. Seiberg, F. Wilczek, and E. Witten.  I am also grateful to
M. Shifman for asking me to write this review, and to T. De Grand and
T. Suzuki for encouraging me to push these ideas forward.  This work
was supported by the National Science Foundation under Grant
PHY-9513835 and by the WM Keck Foundation, and by Department of Energy
grant DE-FG02-95ER40893.




\end{document}